\documentclass[journal,twoside,web]{ieeecolor}
\usepackage{generic}

\usepackage[style=ieee, backend=bibtex, doi=false, url=false, isbn=false, date=year]{biblatex}
\AtBeginBibliography{\footnotesize}
\AtEveryBibitem{\clearlist{language}}
\AtEveryBibitem{ \ifentrytype{misc}{}{\clearfield{note}}}

\usepackage{textcomp}
\usepackage{amsmath}
\usepackage{amssymb,amsfonts, nccmath}
\usepackage{mathrsfs}
\usepackage{mathtools}
\usepackage{ntheorem}
\usepackage{graphicx}
\usepackage{hyperref}
\usepackage{textcomp}
\usepackage{algorithm}
\usepackage{algpseudocode}
\usepackage{caption}
\usepackage{subcaption}
\usepackage{url}

\usepackage[american]{circuitikz}
\usepackage{layouts}

\hypersetup{
	colorlinks=true
}

\newcommand{\R}{\mathbb{R}}
\newcommand{\I}{\mathbf{I}}
\newcommand{\Ex}{\mathfrak{E}}

\newcommand{\1}{\mathbf{1}}

\newcommand{\C}{\mathbb{C}}
\newcommand{\D}{\mathcal{D}}
\newcommand{\N}{\mathbb{N}}

\newcommand{\cN}{\mathcal{N}}
\renewcommand{\j}{\hat{\jmath}}
\renewcommand{\i}{\imath}
\renewcommand{\Re}{\mathrm{Re}}

\newcommand{\Lt}{\mathscr{L}}

\newcommand{\lmin}{\lambda_{\mathrm{min}}}
\newcommand{\lmax}{\lambda_{\mathrm{max}}}

\newcommand{\Lb}{\mathbf{L}}

\newcommand{\Rb}{\mathbf{R}}
\newcommand{\Cb}{\mathbf{C}}

\newcommand{\Gb}{\mathbf{G}}

\newcommand{\m}{\mathrm{m}}

\DeclareMathOperator*{\diag}{diag}

\DeclareMathOperator*{\argmin}{arg\,min}
\renewcommand\d{\mathop{}\!\mathrm{d}}

\theoremseparator{.}
\newtheorem{lemma}{Lemma}
\newtheorem{corollary}{Corollary}
\newtheorem{proposition}{Proposition}
\newtheorem{theorem}{Theorem}
\newtheorem{assumption}{Assumption}
\newtheorem{defn}{Definition}
\newtheorem{problem}{Problem}

\theoremheaderfont{\itshape}
\theoremseparator{:}
\theorembodyfont{\normalfont}
\newtheorem{solution*}{Problem}
\newtheorem{example}{Case Study}
\newtheorem{remark}{Remark}

\def\BibTeX{{\rm B\kern-.05em{\sc i\kern-.025em b}\kern-.08em
    T\kern-.1667em\lower.7ex\hbox{E}\kern-.125emX}}
\begin{document}
\title{Fault Localisation in Infinite-Dimensional Linear Electrical Networks}
\author{Daniel Selvaratnam, \IEEEmembership{Member, IEEE}, Alessio Moreschini, \IEEEmembership{Member, IEEE}, Amritam Das, \IEEEmembership{Member, IEEE}, Thomas Parisini, \IEEEmembership{Fellow, IEEE}, and Henrik Sandberg, \IEEEmembership{Fellow, IEEE}
\thanks{Submitted for review on \today. Supported in part by the Swedish Energy Agency and ERA-Net Smart Energy Systems (project RESili8, grant agreement No 883973), by the Swedish Research Council (Grant 2016-00861), and by the
European Union’s Horizon 2020 Research and Innovation Programme
under Grant 739551 (KIOS CoE).}
\thanks{D. Selvaratnam and H. Sandberg are with the Division of Decision and Control Systems, KTH Royal Institute of Technology, SE-100 44 Stockholm, Sweden. (e-mails: selv@kth.se, hsan@kth.se)}
\thanks{A. Moreschini and T. Parisini are with the Department of Electrical and Electronic Engineering, Imperial College London, SW72AZ London, U.K. T. Parisini is also with the Department of Electronic Systems, Aalborg University, Denmark, and with the Department of Engineering and Architecture, University of Trieste, Italy (e-mails: a.moreschini@imperial.ac.uk, t.parisini@imperial.ac.uk).}
\thanks{A. Das is with the Department of Electrical Engineering, Eindhoven University of Technology, P.O. Box 513, 5600 MB Eindhoven, The Netherlands. (e-mail: am.das@tue.nl)}}

\maketitle

\begin{abstract}
We present a novel fault localisation methodology for linear time-invariant electrical networks with infinite-dimensional edge dynamics and uncertain fault dynamics. The theory accommodates instability and also bounded propagation delays in the network. The goal is to estimate the location of a fault along a given network edge, using sensors positioned arbitrarily throughout the network. Passive faults of unknown impedance are considered, along with stable faults of known impedance. To illustrate the approach, we tackle a significant use-case: a multi-conductor transmission line, with dynamics modelled by the Telegrapher's equation, subject to a line-to-ground fault. Frequency-domain insights are used to reformulate the general fault localisation problem into a non-convex scalar optimisation problem, of which the true fault location is guaranteed to be a global minimiser. Numerical experiments are run to quantify localisation performance over a range of fault resistances.
\end{abstract}


\section{Introduction}
Faults in large-scale physical networks can cause major disruptions to civic life. Think of power-outages, burst water mains, blocked sewerage, or severed internet cables.
Since such are inevitable, rapid fault localisation becomes a matter of pressing practical concern. Major electrical faults are often preceded by smaller transient faults, which, if detected and localised quickly, can be mitigated before catastrophic service disruptions occur~\cite{habib_fault_2022}. Localisation by physical inspection is extremely costly, because even when the faulty network edge is known, a tedious manual search along it may be required. In the power grid, for example, edges are power lines that can stretch for hundreds of kilometres~\cite[Figure 5]{thorslund_swedish_2017}. 
Thus, to facilitate early intervention and possibly avoid manual inspection, a fault localisation methodology that uses measurements taken at fixed nodes in the network is highly desirable.
Nodes are ports where energy and information can be injected or extracted, and edges permit the (potentially lossy) exchange of energy or information between nodes. Flow along the edges is governed by known physical laws, which determine the speed of information propagation. In large-scale networks, this phenomenon introduces a discernible delay between the occurrence of a fault along an edge, and the detection of it at a sensor. This delay implies infinite-dimensional edge dynamics and, in general, necessitates estimation of the fault occurrence time as part of the localisation. Although edge models are usually available, the fault dynamics is never known in practice, due to its unpredictable nature. Therefore, there is a need for a fault localisation technique that accommodates infinite-dimensional edge dynamics, uncertainty in the fault model, and measurements from a limited subset of nodes. This work offers a novel solution for linear time-invariant (LTI) electrical networks.

Together with the localisation algorithm that is presented, our primary contribution is to formulate the problem such that it yields to standard tools of complex analysis and linear algebra, thus making our solution of interest for practitioners. Our attention is focused on scalar fault locations, which presumes prior identification of the faulty edge by other means.
In fact, rather than searching over a Cartesian product space of edges and distances, it is more efficient to first identify the faulty edge and then estimate the fault distance along it. The edge identification step is an instance of the \emph{fault passage indication} problem, for which a number of tools are already available~\cite[Chapter 3]{habib_fault_2022}. We focus instead on the distance estimation step here. 

The network is viewed as an $n$-port~\cite{anderson_network_1973} that is both linear and time-invariant. Analysis is conducted in the frequency domain using non-rational transfer function models, which allow for infinite-dimensional dynamics. Since the partitioning of port signals into inputs and outputs need not reflect the physical causal mechanisms at work~\cite{willems_behavioral_2007}, propagation delays in the network may result in an `output' temporally preceding an `input', thus rendering the network model acausal~\cite[Remark 2]{selvaratnam_frequency-domain_2025}. This can be accounted for if an upper-bound on the delay is known. We permit uncertainty in the fault if the fault is passive~\cite{Brogliato_2020, van2017l2}, meaning that it only stores and dissipates energy, without generating it. \emph{Positive realness}~\cite{anderson_network_1973,kottenstette2014relationships}, which is the frequency-domain characterisation of passivity, can be exploited in the absence of more detailed model information. Two versions of the fault localisation problem are therefore posed. In the first, the fault impedance is known perfectly. In the second, only its passivity is assumed. Each problem is reformulated into a 2-dimensional optimisation problem, over both fault location and time. The problems are then reduced to a single dimension via the construction of a  fault-time estimator. Since they are not convex, a unique solution cannot be guaranteed, but producing a set potential locations is still of immense value for reducing manual inspection time. The novelty of our approach to fault localisation thus lies in its \emph{i)} adoption of classical network theory~\cite{anderson_network_1973} to accommodate a variety of network topologies and sensor locations, \emph{ii)} frequency-domain analysis that allows for delays and infinite-dimensional dynamics, \emph{iii)} reliance on passivity to overcome fault model uncertainty, and \emph{iv)} principled reduction of the problem to scalar optimisation over a compact interval. 

The assumptions made in the problem formulation highlight the mathematical properties crucial to our solution.
To motivate them, and explain their physical significance, a faulted multi-phase transmission line is introduced as a case study. Individual line segments are modelled by the Telegrapher's equation, which is the canonical distributed-parameter (i.e., infinite-dimensional) model of a transmission line~\cite{paul_analysis_2007}. We illustrate, in particular, the construction of the network transfer matrix. To show that it satisfies the assumptions, the recent bounds devised by the authors in~\cite{selvaratnam_frequency-domain_2025} are used, which fully account for line losses, interactions between phases, and asymmetry in the conducting wires, as may occur under fault conditions. A precursor to this multi-phase case study is~\cite{selvaratnam_electrical_2023}, which estimates the length of a single-phase transmission line segment, given voltage and current measurements at one end of the line, assuming both voltage and fault resistance are known at the other. A frequency-domain least-squares estimator is proposed there, and evaluated numerically, but rigorous proofs are not provided. Although the line is also modelled by the Telegrapher's equation, the simplicity of the network structure in \cite{selvaratnam_electrical_2023} allows fault-time estimation to be avoided.

Our work falls within the realm of model-based fault isolation~\cite{ding_model-based_2013,Kinnaert,Parisini-isolation}. Many such studies are limited to a finite set of possible fault scenarios. For example, methods based on electromagnetic time-reversal address a similar problem to ours, but ignore losses in the line, while permitting only a finite set of possible fault locations~\cite{razzaghi_efficient_2013,wang_electromagnetic_2020}. By contrast, we allow for a continuum of fault locations and signatures, fully leveraging the capabilities of distributed-parameter models. In the taxonomy of \cite{dey_robust_2019}, ours is a `late lumping' approach, in which the residual is formed in an infinite-dimensional space, with discretisation only performed at the point of numerical evaluation.
Unlike \cite{demetriou_fault_1996, demetriou_model-based_2002,moura_adaptive_2013,hasan_boundary_2016,aamo_leak_2016, cai_model-based_2016,anfinsen_leak_2022,smyshlyaev_backstepping_2005}, we do not attempt the construction of observers, but instead form the residual in the frequency domain. This obviates the need for such sophisticated machinery as PDE backstepping and integral transformation methods~\cite{smyshlyaev_backstepping_2005,fischer_fault_2022,ascencio_adaptive_2016,krstic_boundary_2008,10767284}, allowing us to rely instead on classical linear analysis. Works such as \cite{demetriou_online_2002,smyshlyaev_backstepping_2005,Ascencio,dey_robust_2019,demetriou_fault_2022} restrict attention to parabolic PDEs, which have an infinite speed of information propagation, whereas our framework accommodates propagation delays, thus admitting hyperbolic systems as well. Like~\cite{dey_robust_2019}, we consider uncertainty in the fault model, but reserve uncertainty in our network model for future work. 
The focus of this brief review has been on model-based fault localisation methods for distributed parameter systems. Comprehensive reviews of electrical fault localisation specifically can be found in~\cite{habib_fault_2022, parmar_fault_2015}.

After establishing notation and basic definitions below, the remainder of this paper proceeds as follows. The fault localisation problem is formulated in Section~\ref{sec:prob} for both known and unknown fault impedances, and the transmission line case study introduced. Its transfer matrix is constructed in Section~\ref{sec:modelling}, before the localisation problem is solved in Section~\ref{sec:soln}. Section~\ref{sec:sim} then evaluates estimation performance on the simulated case study, before concluding remarks are offered in Section~\ref{sec:conc}.

\subsection{Nomenclature}
\subsubsection{Sets and numbers} \label{sec:sets} The subset relation is denoted by $\subseteq$, and its strict version by $\subset$. 
Let $\C$ denote the complex numbers, $\j \in \C$ the imaginary unit, $\R$ the reals, $\j \R$ the imaginary axis, $\mathbb{Z}$ the integers, and $\N$ the naturals with $0\in\N$. Given $\alpha \in \R$, define $\C^+_\alpha:= \{ s \in \C \mid \Re(s)>\alpha\}$, with $\C^+:= \C^+_0$. Let $\overline{S}$ denote the closure of $S \subseteq \C$ relative to $\C$, and ${S^+:= S \cap \C^+}$.  For $x \in \R$, \begin{equation} x^+:= \max\{0,x\}.\label{eq:xplus} \end{equation}
\subsubsection{Linear algebra} Given a matrix $A \in \C^{m \times n}$,  its $k$th singular value is $\sigma_k(A)$, its conjugate transpose $A^*$, and its Moore-Penrose pseudoinverse $A^\dagger$. If it is Hermitian, then $A \succ 0$ asserts positive definiteness, and $A \succeq 0$ positive semi-definiteness. Moreover, its maximum and minimum eigenvalues are $\lmax(A)$ and $\lmin(A)$, respectively. Denote the $n \times n$ identity matrix by $\I_n$. The Frobenius matrix norm is $\| \cdot \|_F$.  Let $\| \cdot \|_p$ return the $p$-norm of a vector in $\C^n$, and for a matrix, the corresponding $p,p$ induced norm. To streamline notation, $\| \cdot \| := \| \cdot \|_2$.
\subsubsection{Functions} 
The unit step is
\begin{align} \1(t) := \begin{cases}
		0,& \text{if } t < 0, \\
		1,& \text{if } t \geq 0.
	\end{cases}  \label{eq:shortcuts}
\end{align}
Given a function ${f:[0,\infty) \to \R^n}$ denoted by a lower-case letter, the corresponding upper-case letter is reserved for its Laplace transform
$$ F(s) := \Lt[f](s) : = \int_0^\infty f(t) e^{-st} \d t. $$
Conversely, $f(t) = \Lt^{-1}[F](t) .$ 

\begin{defn}[Exponential order signals]
	Let $\Ex(\alpha):= \bigcup_{n=1}^\infty \Ex^n(\alpha)$, where $\Ex^n(\alpha)$ is the set of locally integrable functions $f:[0,\infty) \to \R^n$ such that
    $$ \exists M,T \geq 0,\ \forall t \geq T,\ \|f(t) \| \leq Me^{\alpha t}.$$ 
\end{defn}
Above, $\Ex(\alpha)$ is the set of locally integrable signals of exponential order $\alpha$, all of which have Laplace transforms that are analytic on $\C^+_\alpha$~\cite{schiff_laplace_1999}.
\smallskip
\begin{defn}[H-infinity] \label{def:Hinf}
	Let $\mathcal{H}_\infty := \bigcup_{m=1}^\infty \bigcup_{n=1}^\infty \mathcal{H}^{m \times n}_\infty,$ where $\mathcal{H}^{m \times n}_\infty$ is the space of functions $H$ such that: \begin{enumerate}
		\item $H:S \to \C^{m \times n}$ for some $S \subseteq \C$;
		\item $H$ is analytic on $\C^+ \subseteq S$;
		\item $ \sup \{ \| H(s) \| \mid s \in \C^+ \} < \infty$. \label{cl:Hinf}
	\end{enumerate}
\end{defn}
\smallskip
The space $\mathcal{H}_\infty$ is the set of transfer functions of causal $\mathcal{L}_2$-stable LTI systems~\cite[Chapter 3.4.3]{dullerud_course_2000}.

\section{Problem Formulation} \label{sec:prob}
\subsection{Network modelling}
\begin{figure}[t]
	\centering
	\begin{circuitikz}
		\tikzset{quad/.style={draw, thick, minimum height=4cm, minimum width=4cm}}
		\node[quad] (A) at (0,0) {$\mathcal{N}$};
		\draw 
		(A.west)++(0,1) to[short,-o, i<_= $i_1$] ++(-0.5,0) to[short] ++(-1,0) to[V] ++(0,-2) to[short] ++(1,0) to[short,o-] ++(0.5,0.)
		(A.west)++(-0.5,1) to[open, v_= $v_1$] ++(0,-2)
		
		(A.east)++(0,1) to[short,-o, i<^= $i_2$] ++(0.5,0) to[short] ++(1,0) to[I,invert] ++(0,-2) to[short] ++(-1,0) to[short,o-] ++(-0.5,0.)
		(A.east)++(0.5,1) to[open, v_= $v_2$] ++(0,-2)
		
		(A.north)++(-1,0) to[short,-o, i<^= $i_3$] ++(0,1) to[ammeter] ++(0,1.5) to[twoport, t=$(\cdot)$] ++(2,0) to[short] ++(0,-1.5) to[short,o-] ++(0,-1)
		(A.north)++(-1,1) to[voltmeter, v_= $v_3$] ++(2,0)
		
		(A.south)++(-1,0) to[short,-o, i<_= $i_4$] ++(0,-0.5) to[switch, l_={$t_f$}] ++(0,-1) to[twoport, t=$\Phi(s)$, v=$v_f$] ++(2,0) to[short] ++(0,1) to[short,o-] ++(0,0.5)
		(A.south)++(-1,-0.5) to[open, v_= $v_4$] ++(2,0)
		;
	\end{circuitikz}
	\caption{Four-port LTI Network $\cN$ with fault. \label{fig:faultNetwork}}
\end{figure}
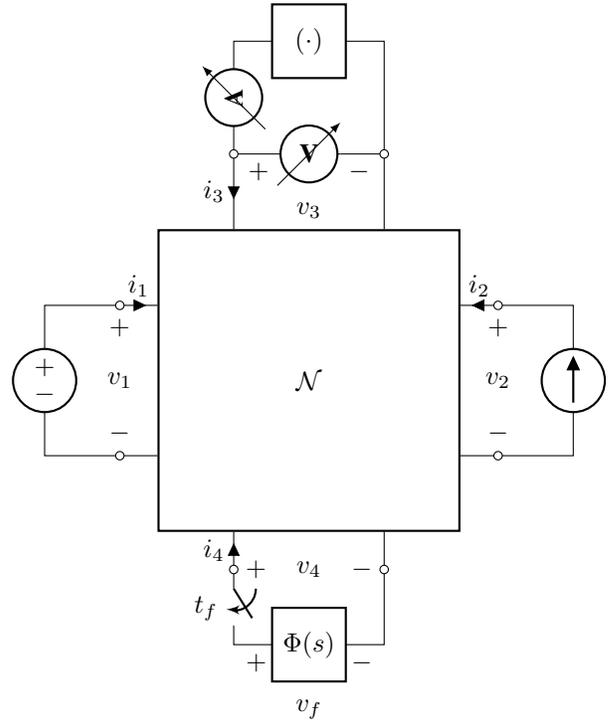

Consider the 4-port network $\cN$ depicted in Figure~\ref{fig:faultNetwork}. Each port is multi-dimensional, and can be viewed as a collection of scalar ports that are grouped according to the information available at each port. The first port is connected to a known voltage source, of dimension $n_1$, and the second to a known $n_2$-dimensional current source. Both voltage and current measurements are taken at the third port, which is of dimension $n_3$, and connected to an arbitrary external network. The fourth port, of dimension $n_4$, is connected to a fault, which is modelled as a switch in series with an LTI network of impedance $\Phi(s) \in \C^{n_4 \times n_4}$. For each ${k \in \{1,...,4\}}$, $v_k(t) \in \R^{n_k}$ and $i_k(t) \in \R^{n_k}$ denote, respectively, the voltage across and current into the $k$th port of $\cN$ at time $t \geq 0$. 

It is worth noting that the sources at ports 1 and 2 are intended to be typical, but not restrictive. For example, the voltage source may be an artificial one, arising from the generalised Thevenin equivalent~\cite[Section 7.3.1]{paul_analysis_2007} of a fault-free portion of the network. Similarly, the current source may arise from the generalised Norton equivalent~\cite[Section 7.3.2]{paul_analysis_2007}. The crucial point is that $v_1$ and $i_2$, along with $v_3$ and $i_3$, are {\it known signals}.

\smallskip
\begin{assumption}[Well-posedness] \label{ass:wellposed}
	Let $\alpha \in \R$. For every ${k \in \{1,...,4\}}$, the signals $i_k,v_k \in \Ex^{n_k}(\alpha)$, where $n_k \in \N^+$ .
\end{assumption}
\smallskip
 Assumption~\ref{ass:wellposed} ensures that all port signals have well-defined Laplace transforms that share a common region of convergence, $\C^+_\alpha$. Since $\alpha $ is permitted to be positive, exponentially growing port signals, and thus unstable networks, are accommodated. 
 
\begin{figure}[t]
	\begin{subfigure}{\columnwidth}
		\centering
		\begin{circuitikz}[scale=0.75]
			\draw (0,0) node[ground]{} to (0,1.5) to[V,v_=$v_1$, invert, o-o] (0,3.5) to[short,i=$i_1$](0,5) to[transmission line, label = $D - \ell$, bipoles/tline/width=1] (4.5,5) to[transmission line, label=$\ell$,bipoles/tline/width=1] (9,5) to[rmeter,t=A, i^<=$i_3$] (9,2.5) to[twoport, t = $(\cdot)$, o-o] (9,0) node[ground]{}; 
			\draw (9,2.5) to (6.7,2.5) to[rmeter,t=V,l=$v_3$] (6.7,0) to (9,0);
			\draw (4.5,5) to[short,i^<=$i_4$,-o] (4.5,3) to[switch,label = {$t_f$}](4.5,2) to[twoport, t = $\Phi(s)$, o-o] (4.5,0) node[ground]{};
			\draw [dashed] (4.5,3) to (3,3) to[open, v = $v_4$,o-o] (3,0) to (4.5,0);
		\end{circuitikz}
		\caption{Faulted transmission line, connected to voltage source and sensors.\label{fig:faultDiagram}}
	\end{subfigure}
	\vspace{1em}
	\begin{subfigure}{\columnwidth}
		\vspace{2em}
		\centering
		\begin{circuitikz}[scale=0.9]
			\draw 
			(0,1) to[short,-o,i<=$i_4$] (0,0.5) to[open, v = $v_4$] (0,-0.5) to[short,o-] (0,-1)
			node[fourport,t={$\Upsilon(s{,}D-\ell)$}] at (-2.5,0) (c) {}
			(c.port3) to[short,i<=$i_4^+$] ++(1,0) coordinate (T); 
			\draw (T) |- (0,1);
			\draw (c.port2) --++(1,0) |- (0,-1);
			\draw node[fourport,t={$\Upsilon(s{,}\ell)$}] at (2.5,0) (d) {}
			(d.port4) to[short,i_<=$i_4^-$] ++(-1,0) coordinate (S); 
			;
			\draw (S) |- (0,1);
			\draw (d.port1) --++(-1,0) |- (0,-1);
			\draw (c.port4) to[short,i_<=$i_1$,-o] ++(-1,0) coordinate (A)
			(c.port1) to[short,-o] ++(-1,0) to[open,v<=$v_1$] (A);
			
			\draw (d.port3) to[short,i<=$i_3$,-o] ++(1,0) coordinate (B)
			(d.port2) to[short,-o] ++(1,0) to[open,v<=$v_3$] (B);
		\end{circuitikz}
		\caption{Three-port representation of faulty line.\label{fig:faultDiagramPort}}
	\end{subfigure}
	\caption{Diagram of Case Study \ref{ex:tline}, an illustrative use-case. \label{fig:faultExample}}
\end{figure}
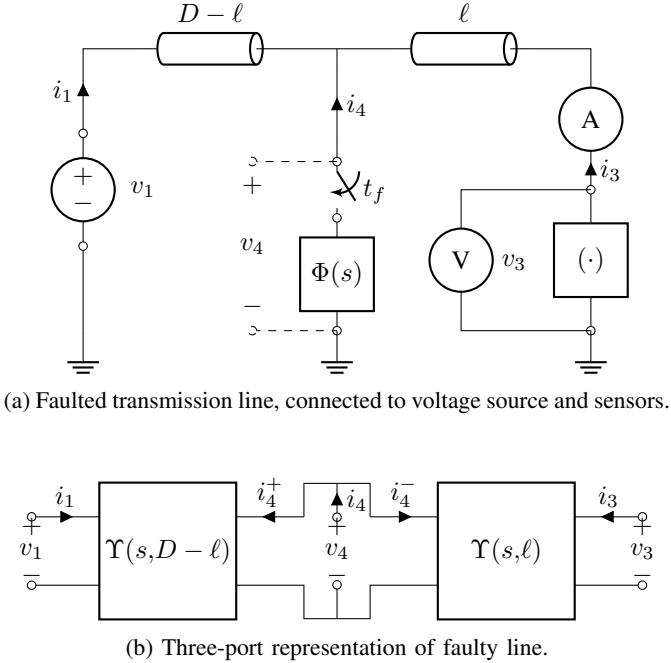

 The next assumption prescribes the relation between the port signals and the fault location. Below, $\D$ represents the search domain, $d$ a scalar distance coordinate, and $\ell$ the true fault location. Implicit in the assumption is a local coordinate system for the network that assigns a unique position $d=\ell$ to the fault. Since $\cN$ can have many branches, for $\ell$ to be scalar, the faulty edge must be known $\emph{a priori}$. Then, $d$ is simply the distance along it from a given endpoint. 
 \begin{assumption}[Network admittance] \label{ass:model} 
 	Let $\D \subset \R$ be a compact interval. Define ${i(t):= (i_1(t), \hdots, i_4(t)) \in \R^N}$ and ${v(t):= (v_1(t), \hdots, v_4(t) ) \in \R^N}$, where $N:= \sum_{k=1}^4 n_k$. The transfer function ${Y:\C^+_\alpha \times \D \to \C^{N \times N}}$is such that
 	\begin{enumerate}
 		\item for every $d \in \D$, the map $ s \mapsto Y(s;d)$ is analytic; \label{cl:analytic}
 		\item  at the fault location $\ell \in \D$, \label{cl:attruth} $$\forall s \in \C^+_\alpha,\ I(s) = Y(s;\ell)V(s);$$
 		\item given known parameters $\beta > \max\{0,\alpha\}$ and $\tau \geq 0$, there exists $M \geq 1$ such that
 		\begin{equation} \forall s \in \overline{\C^+_\beta},\ \ \forall d \in \D,\ \| Y(s;d) \| \leq  Me^{\tau \Re(s) }.  \label{eq:normbound} \end{equation}
 	\end{enumerate} 
 \end{assumption}
 \smallskip
 Assumption \ref{ass:model} requires the network $\cN$ to possess an admittance matrix $Y(s;d)$, parametrised by the distance coordinate $d$. Clause \ref{cl:analytic} is necessary for $Y$ to be a transfer function. The second clause demands that the admittance relation $I(s) = Y(s;\ell) V(\ell)$ hold at the true fault location. The third clause permits the admittance matrix to be acausal, so that propagation delays between the fault port and the measured signals can be accounted for, with $\tau$ being a known bound on the maximum delay~\cite[Remark 2]{selvaratnam_frequency-domain_2025}. The parameter $\beta$ is chosen such that $ \overline{\C^+_\beta} \subset \C^+ \cap \C^+_\alpha$. Rationality of $Y(s;d)$ is not required, so the assumption admits infinite-dimensional dynamics.
\smallskip
\begin{remark} \label{rem:coordinate}
	Since $ \D \subset \R$, Assumption \ref{ass:model} relies on prior determination of the faulty edge, a problem not addressed here. Without this knowledge, the search domain expands to a subset of $\mathcal{E} \times \D$, where $\mathcal{E}$ is the edge set. Much of the subsequent analysis remains valid, but finding a solution may be computationally intractable. It is more efficient to first identify the faulty edge by other means.
\end{remark}
\smallskip
Assumption \ref{ass:model} does not require continuity of $Y$ with respect to $d$, as it is not essential for much of the theory that follows. It is, however, stated as a separate assumption below, because it simplifies the numerical search for the fault location, and can be reasonably expected on physical grounds.
\smallskip
\begin{assumption}[Continuity]\label{ass:continuous}
	The network admittance matrix $Y:\C^+_\alpha \times \D \to \C^{N \times N}$ is continuous.
\end{assumption}
\smallskip
The construction of an admittance matrix, for a given network, with all the above properties is the first step to solving the fault localisation problem.  Although an automated construction procedure is beyond the scope of this work, the construction is illustrated by means of a case study, for the reader's convenience.
\smallskip
\begin{example}[Transmission line] \label{ex:tline}
	A particular instance of the network $\cN$ in Figure~\ref{fig:faultNetwork} is depicted in Figure~\ref{fig:faultExample}.
	Figure~\ref{fig:faultDiagram} shows a transmission line of length $L>0$, with a voltage source at one end, voltage and current sensors at the other, and a line-to-ground fault in-between at distance $\ell > 0$ from the sensors. Figure~\ref{fig:faultDiagramPort} omits the external connections in order to emphasise the 3-port structure of the faulty line. Although only a single phase is drawn for simplicity, assume the lines are $n$-phase, together with the external voltage source and sensors. Thus, $n_1 = n_3 = n$. Without loss of generality, it can also be assumed that $n_4=n$, even if the fault does not affect all phases. Observe that Port 2 is missing. Since $n_2 = 0$ would violate Assumption \ref{ass:wellposed}, a dummy port of dimension $n_2 = 1$ can be added in its place, which simply performs the identity map $i_2 = v_2$ without interacting with the other ports (i.e., a floating unit resistor).
\end{example}

\smallskip
In general, not every LTI electrical network has an admittance matrix. For example, short circuits and ideal transformers do not~\cite[Chapter 2.4]{anderson_network_1973}. If necessary, the localisation procedure of Section~\ref{sec:soln} is readily adapted to other transfer function models, such as impedance, hybrid, chain or scattering matrices. An admittance model is adopted here, because the focus is on transmission lines: Line branches are parallel interconnections,  for which admittance matrix composition takes a particularly simple form. 
Section~\ref{sec:modelling} applies these composition rules to construct an admittance matrix for Case Study \ref{ex:tline}, before the general localisation problem is tackled in Section~\ref{sec:soln}.

\subsection{Fault localisation}
The fault model is specified next. Below, $v_f(t) \in \R^{n_4}$ denotes the voltage across the fault, as shown in Figure \ref{fig:faultNetwork}. The presence of the switch distinguishes it from the port voltage $v_4(t)$.
\smallskip
\begin{assumption}[Fault model] \label{ass:faultModel}
	The fault time $t_f \in (0,T_f)$ and fault impedance $\Phi \in \mathcal{H}^{n_4 \times n_4}_\infty$ are such that
	\begin{subequations}
		\begin{align}
			&\forall t \in [0,t_f),\ &&i_4(t) =0,\label{eq:zerocurrent}\\
			&\forall t \geq 0,\ &&v_f(t) = \1(t - t_f) v_4(t), \label{eq:switch}\\
			&\forall s \in \C^+ \cap \C^+_\alpha,\ &&V_f(s) =  - \Phi(s) I_4(s). \label{eq:Ohms}
		\end{align}
	\end{subequations}
\end{assumption}
\smallskip
Time $T_f > 0$ is simply a known upper-bound on the time $t_f$, at which the fault occurs. Equations \eqref{eq:zerocurrent} and \eqref{eq:switch} assert that both fault voltage and current are zero prior to the switch closing at time $t_f$, after which the fault voltage equals the port voltage $v_4$. Equation \eqref{eq:Ohms} requires the fault voltage to also satisfy the impedance relation $V_f(s) = \Phi(s) I_f(s)$ in the frequency domain, where $i_f = -i_4$ is the current into the fault. Since $\Phi \in \mathcal{H}_\infty$, the fault impedance is assumed to be stable and causal, but {\it need not be finite-dimensional}.

 The fault localisation problem for electrical networks can now be stated precisely.
\smallskip
\begin{problem}[Known fault impedance] \label{prob:knownPhi}
	Under Assumptions~\ref{ass:wellposed}--\ref{ass:faultModel}, estimate the fault location $\ell \in \D$, given knowledge of the fault impedance $\Phi$ and port signals $v_1,i_2,v_3,i_3$, without knowing the fault time $t_f \in (0,T_f)$.
\end{problem}
\smallskip
Problem~\ref{prob:knownPhi} assumes the fault impedance is known. When it is unknown, location information can still be extracted from the measurements under the additional assumption that the fault is \emph{passive}, which means it can only store or dissipate energy, without generating it.
\smallskip
\begin{assumption}[Passivity] \label{ass:passive}
	The fault impedance matrix ${\Phi:\C^+ \to \C^{n_4 \times n_4}}$ is positive real. That is, $\Phi$ is analytic, and both the following hold: 
	\begin{align*}
		\forall s \in \R^+,\ &\Phi(s) \in \R, \\
		\forall s \in \C^+,\  & \Phi(s)^* + \Phi(s) \succeq 0.
	\end{align*}
\end{assumption}
\smallskip
Recall that positive-realness is the frequency-domain characterisation of passivity~\cite{kottenstette2014relationships}. As posed below in Problem~\ref{prob:passivePhi}, not knowing the fault impedance leads to a more challenging localisation problem, but one that better reflects the situation in practice.
\smallskip
\begin{problem}[Unknown passive fault] \label{prob:passivePhi}
	Under Assumptions~\ref{ass:wellposed}--\ref{ass:passive}, estimate the fault location $\ell \in \D$, given the knowledge of the port signals $v_1,i_2,v_3$ and $i_3$, but without knowing the fault time $t_f \in (0,T_f)$ or impedance $\Phi$.  
\end{problem}

\section{Admittance modelling} \label{sec:modelling}

In this section, we construct the admittance matrix for Case Study~\ref{ex:tline}. Although, the topology is assumed to have a simple form for the sake of expositional clarity, the same procedure can be applied to networks with more complicated topologies. The first step is to construct an admittance matrix $\Upsilon(s,d) \in \C^{2n \times 2n}$ for the transmission line element in Figure~\ref{fig:faultExample}. The second step deals with composing the admittances of each element to form the admittance matrix of the whole network. We show that the resulting admittance satisfies Assumptions~\ref{ass:model}--\ref{ass:continuous}.

\subsection{Transmission line element}

The transmission line element represents a single unbroken transmission line segment of finite length. Several models for this are available, both finite- and infinite-dimensional. The Telegrapher's equation is chosen here, because it models the finite speed of information propagation along the line. It is also the canonical distributed parameter model~\cite{paul_analysis_2007}. In \cite{selvaratnam_frequency-domain_2025}, the line segment admittance matrix is constructed from its ABCD matrix, which is in turn derived directly from the Telegrapher's equation. Let the symmetric matrices $\Lb,\Cb,\Rb,\Gb \in \R^{n \times n}$ denote, respectively, the inductance, capacitance, resistance and conductance matrices of the line, with $\Lb,\Cb \succ 0$. Then, the ABCD matrix for a line segment of length $d$ reads
$$ \Xi(s,d) := e^{d\begin{bsmallmatrix}
		0 & \Lb s + \Rb \\ \Cb s + \Gb & 0
\end{bsmallmatrix}} = \begin{bmatrix}
	A_d(s) & B_d(s) \\ C_d(s) & D_d(s)
\end{bmatrix},$$
where explicit expressions for the $n \times n$ A, B, C, and D blocks are provided in \cite[Theorem 1]{selvaratnam_frequency-domain_2025}. The admittance matrix is given in terms of them by
\begin{equation} \Upsilon(s,d) = \begin{bmatrix}
	D_d(s)B_d(s)^{-1} & -B_d(s)^{-1} \\
	-B_d(s)^{-1} & D_d(s)B_d(s)^{-1}
\end{bmatrix}, \label{eq:Upsilon} \end{equation}
where $\Upsilon : \C^+_\alpha \times \R^+ \to \C^{2n \times 2n}$ for
$$  \alpha= \medmath{ -\min\left\{\tfrac{\lmin(\Gb)}{\lmax(\Cb)},\tfrac{\lmin(\Gb)}{\lmin(\Cb)},\tfrac{\lmin(\Rb)}{\lmax(\Lb)},\tfrac{\lmin(\Rb)}{\lmin(\Lb)} \right\}.} \label{eq:alpha} $$
Continuity of $\Upsilon$ is established in \cite[Theorem 3]{selvaratnam_frequency-domain_2025}, along with analyticity in $s$ for every $d$.
\subsubsection{Growth bounds} \label{sec:growth}

A constant $\nu>0$, which lower-bounds the line propagation speed, can be computed from the line parameters $\Lb,\Cb,\Rb$ and $\Gb$ using \cite[(21b) and Lemma 4]{selvaratnam_frequency-domain_2025}. Assumption \ref{ass:model} stipulates that $\D$ be a compact interval, and if $\inf \D > 0$, then Clause~3 of \cite[Theorem 3]{selvaratnam_frequency-domain_2025} implies the following growth bound: for any choice of ${\beta \in [0,\infty) \cap (\alpha,\infty)}$, there exists $M_1 > 0$ such that
\begin{equation}
	\forall s \in \overline{\C^+_\beta}\ \forall d \in \D,\	\| \Upsilon(s,d) \| \leq M_1 e^{\frac{nd}{\nu} \Re(s)} \, .\label{eq:Me1}
\end{equation}
A similar bound on the ABCD matrix follows from \cite[Theorem 2]{selvaratnam_frequency-domain_2025}: there exists $M_2 > 0$ such that
\begin{equation}
	\forall s \in \overline{\C^+} \ \forall d \in \D,\	\| \Xi(s,d) \| \leq M_2 e^{\frac{d}{\nu} \Re(s)} \, . \label{eq:Me2}
\end{equation}

\subsection{Admittance matrix composition} \label{sec:composition}
It now remains to compose the admittances for each line segment to form the admittance matrix for the whole network. The procedure illustrated below is based on well-known composition rules for $n$-port networks~\cite[Chapter 2.5]{anderson_network_1973}, which are readily applied ``by hand" to other networks of interest. 

Let us refer again to Figure~\ref{fig:faultExample}. By definition of the admittance matrix, the relation between voltages and currents downstream of the fault is
\begin{equation}
	\begin{bmatrix}
		I_3(s) \\ I_4^-(s)
	\end{bmatrix} = \begin{bmatrix} \Upsilon_1(s,\ell) & \Upsilon_2(s,\ell) \\ \Upsilon_3(s,\ell) & \Upsilon_4(s,\ell) \end{bmatrix}\begin{bmatrix}
		V_3(s) \\ V_4(s)
	\end{bmatrix}, \label{eq:downstream}
\end{equation}
where the line admittance matrix $\Upsilon(s,\ell) = \begin{bsmallmatrix}
	\Upsilon_1(s,\ell) & \Upsilon_2(s,\ell) \\ \Upsilon_3(s,\ell) & \Upsilon_4(s,\ell)
\end{bsmallmatrix}$ is partitioned into four $n \times n$ blocks, as per \eqref{eq:Upsilon}. Similarly, for the upstream portion, we get
\begin{equation}
	\begin{bmatrix}
		I_1(s) \\ I_4^+(s)
	\end{bmatrix} = \begin{bmatrix} \Upsilon_1(s,L-\ell) & \Upsilon_2(s,L-\ell) \\ \Upsilon_3(s,L-\ell) & \Upsilon_4(s,L-\ell) \end{bmatrix} \begin{bmatrix}
		V_1(s) \\ V_4(s)
	\end{bmatrix}. \label{eq:upstream}
\end{equation}
Since $I_4 = I_4^+ + I_4^-$ by Kirchkoff's Current Law, we obtain
\begin{multline}	\begin{bmatrix}
		I_1(s) \\ I_3(s) \\ I_4(s)
	\end{bmatrix} = \Bigg( \begin{bmatrix} \Upsilon_1(s,L-\ell)  & 0 & \Upsilon_2(s,L-\ell)\\ 
		0 & 0 & 0 \\
		\Upsilon_3(s,L-\ell) & 0 & \Upsilon_4(s,L-\ell)
	\end{bmatrix} \\  
	+  \begin{bmatrix} 0 & 0 & 0 \\ 
		0 & \Upsilon_1(s,\ell)  & \Upsilon_2(s,\ell)  \\
		0& \Upsilon_3(s,\ell) &  \Upsilon_4(s,\ell) 
	\end{bmatrix} \Bigg)\begin{bmatrix}
		V_1(s) \\ V_3(s) \\ V_4(s)
	\end{bmatrix}.  \label{eq:attruth}\end{multline}
Therefore, the admittance matrix for a branched line is obtained by adding the admittance matrices of each line segment, after first lifting them to a common product space containing all port voltages and currents. To account for the dummy port mentioned in Case Study \ref{ex:tline}, an additional row and column must also be inserted. Then, the desired distance-parameterised admittance matrix takes on the form
\begin{equation*} \medmath{
		Y(s;d) = \begin{bmatrix} \Upsilon_1(s,L-d) & 0 &  0 & \Upsilon_2(s,L-d) \\ 
			0 & 1 & 0 & 0 \\
			0 & 0 & \Upsilon_1(s,d)  & \Upsilon_2(s,d) \\
			\Upsilon_3(s,L-d) & 0 & \Upsilon_3(s,d)  & \Upsilon_4(s,L-d)  + \Upsilon_4(s,d) 
		\end{bmatrix},}  \end{equation*}
where the arbitrary distance $d$ has been substituted for the true location $\ell$.

Recall the assumption $n_4 = n$ in Case Study \ref{ex:tline}. If the fault does not actually affect all phases, further modification is required:
Any unfaulted components of $i_4$ must be set to zero, because there is no flow of current through to ground on those phases. Accordingly, 
set $ \Pi:= \diag(p_1, \hdots ,p_n) \in \{0,1\}^{n \times n}$, where $p_k = 1$ if and only if the $k$th phase is faulted. Then, $Y(s;d) = $
\begin{equation*} \medmath{
		\begin{bmatrix} \Upsilon_1(s,L-d) & 0 &  0 & \Upsilon_2(s,L-d) \\ 
			0 & 1 & 0 & 0 \\
			0 & 0 & \Upsilon_1(s,d)  & \Upsilon_2(s,d) \\
			\Pi \Upsilon_3(s,L-d) & 0 & \Pi \Upsilon_3(s,d)  & \Pi \Upsilon_4(s,L-d)  + \Pi \Upsilon_4(s,d)
		\end{bmatrix}}  \end{equation*}
is the final result, where owing to~\eqref{eq:Upsilon},
\begin{align*}
	&\Upsilon_1(s,d) =  \Upsilon_4(s,d)  = D_d(s) B_d(s)^{-1}, \\
	&\Upsilon_3(s,d) = \Upsilon_2(s,d)  = - B_d(s)^{-1}.
\end{align*}
Clause \ref{cl:attruth} of Assumption \ref{ass:model} has thus been satisfied by construction. Moreover, the composition operation clearly preserves the continuity and analyticity properties of $\Upsilon$ established in \cite[Theorem 3]{selvaratnam_frequency-domain_2025}.
\subsubsection{Growth bounds}
 The domain of $Y$ must be chosen carefully to satisfy Clause~\ref{cl:Hinf}. Since $\Upsilon(s,d)$ is not defined at $d = 0$, the search domain $\D$ cannot be $[0,L]$. 
Instead, we set the search domain to $\D := [\delta, L - \delta] $, for some arbitrarily small $\delta \in (0, \min\{\ell, L - \ell\}]$. The bound~\eqref{eq:Me1} then applies to $\Upsilon$, thus giving
\begin{align}
	\| Y(s;d) \| & \leq \| \Upsilon(s,d) \| + \| \Upsilon(s,L-d) \| + 1 \nonumber \\
	& \leq M_1 e^{\frac{nd}{\nu} \Re(s)} + M_1 e^{\frac{n(L-d)}{\nu} \Re(s)} + 1 \nonumber \\
	& \leq 2M_1 e^{\frac{nL}{\nu} \Re(s)}  + 1 \nonumber \\
	& \leq (2M_1 + 1) e^{\frac{nL}{\nu} \Re(s)} \, ,\label{eq:Ybound}
\end{align}
for all $s \in \overline{\C^+_\beta}$ and $d \in \D$. The parameter $\beta$ can be chosen arbitrarily, as long as $\beta > \alpha$ and $\beta \geq 0$.
Hence, the admittance $Y: \C^+_\alpha \times \D \to \C^{(3n + 1) \times (3n + 1)} $ satisfies Assumptions~\ref{ass:model} and~\ref{ass:continuous}.

\section{Localisation algorithm} \label{sec:soln}

Having demonstrated the construction of an admittance matrix that satisfies Assumptions~\ref{ass:model} and~\ref{ass:continuous}, this section proposes solutions to Problems~\ref{prob:knownPhi} and~\ref{prob:passivePhi}. In general, a set of candidate solutions to each problem is generated, because a unique estimate of the fault location $\ell$ cannot be guaranteed. These sets are constructed in two steps. By Clause~\ref{cl:attruth} of Assumption~\ref{ass:model}, $I(s) = Y(s;\ell) V(s) $, where only $v_1,i_2,v_3,i_3$ are known signals. The first step is to estimate the fault-port signals $v_4,i_4$ from the known signals, as a functions of the location coordinate $d$. Here, unique signal estimates are obtained. The next step is to formulate an optimisation problem that estimates $\ell$ from the signal estimates by searching over $d \in \D$. This may yield a set of candidate solutions, because the problem is non-convex.

\subsection{Algebraic analysis}

To estimate $v_4, i_4$, the temporal aspect of the problem can be initially ignored, focusing instead on the structure of the linear equation $ I(s) = Y(s;\ell) V(s) $. Thus, for the remainder of this subsection, $Y$ simply denotes a square complex matrix composed of blocks $Y_{jk}$. To distinguish them from matrices, lower-case $i_k$ and $v_k$ denote complex vectors. 
\smallskip
\begin{problem}[Port estimation] \label{prob:port}
	Let $n_k \in \N^+$ and $ Y_{jk} \in \C^{n_j \times n_k}$ for each $j,k \in \{1,...,4\}$. Given known $v_1 \in \C^{n_1}$, $i_2 \in \C^{n_2}$ and $v_3,i_3 \in \C^{n_3}$, find all $i_1 \in \C^{n_1}$, $v_2 \in \C^{n_2}$, and  $v_4,i_4 \in \C^{n_4}$ that together satisfy
	\begin{equation}
		\begin{bmatrix}
			Y_{11} & Y_{12} & Y_{13} & Y_{14} \\
			Y_{21} & Y_{22} & Y_{23} & Y_{24} \\
			Y_{31} & Y_{32} & Y_{33} & Y_{34} \\
			Y_{41} & Y_{42} & Y_{43} & Y_{44}
		\end{bmatrix} \begin{bmatrix}
			v_1 \\ v_2 \\ v_3 \\ v_4
		\end{bmatrix} = \begin{bmatrix}
			i_1 \\ i_2 \\ i_3 \\ i_4
		\end{bmatrix}. \label{eq:Yvi}
	\end{equation}
\end{problem}
\smallskip
Although Problem~\ref{prob:port} is stated in general terms, for our purposes, the unknown $v_2,i_1$ are only of interest insofar as they facilitate estimation of $v_4,i_4$.
\smallskip
\begin{lemma} \label{lem:rearrange}
	The tuple $(i_1,v_2,v_4,i_4) $ solves Problem~\ref{prob:port} if and only if
	\begin{multline}
\begin{bmatrix}
			\I_{n_1} &- Y_{12} & -Y_{14} & 0 \\
			0 & -Y_{22} & -Y_{24} & 0 \\
			0& -Y_{32} & -Y_{34} & 0 \\
			0 & -Y_{42} & -Y_{44} & \I_{n_4}
		\end{bmatrix}  \begin{bmatrix}
			i_1 \\ v_2 \\ v_4 \\ i_4
		\end{bmatrix} \\ = \begin{bmatrix}
			Y_{11} & 0  & Y_{13} & 0 \\
			Y_{21} & -\I_{n_2}  & Y_{23} & 0 \\
			Y_{31} & 0 & Y_{33} & -\I_{n_3} \\
			Y_{41} & 0 & Y_{43} & 0
		\end{bmatrix} \begin{bmatrix}
			v_1 \\ i_2 \\ v_3 \\ i_3
		\end{bmatrix}.\label{eq:YiYv}
	\end{multline}
\end{lemma}
\proof{ Expanding \eqref{eq:Yvi} gives
	\begin{multline}
		\begin{bmatrix}
			Y_{11} \\ Y_{21} \\ Y_{31} \\ Y_{41}
		\end{bmatrix} v_1 + 
		\begin{bmatrix}
			Y_{12} \\ Y_{22} \\ Y_{32} \\ Y_{42} 
		\end{bmatrix} v_2 + 
		\begin{bmatrix}
			Y_{13} \\ Y_{23} \\ Y_{33} \\ Y_{43} 
		\end{bmatrix} v_3 +
		\begin{bmatrix}
			Y_{14} \\ Y_{24} \\ Y_{34} \\ Y_{44}
		\end{bmatrix} v_4  \\ = 
		\begin{bmatrix}
			\I_{n_1} \\ 0 \\ 0 \\ 0
		\end{bmatrix} i_1 + 
		\begin{bmatrix}
			0 \\ \I_{n_2} \\ 0 \\ 0
		\end{bmatrix} i_2 + 
		\begin{bmatrix}
			0 \\ 0 \\ \I_{n_3} \\ 0
		\end{bmatrix} i_3 +
		\begin{bmatrix}
			0 \\ 0 \\ 0 \\  \I_{n_4}
		\end{bmatrix} i_4,
	\end{multline}
	which can be rearranged to yield \eqref{eq:YiYv}. \hfill\QED}
\smallskip

If the LHS of \eqref{eq:YiYv} can be inverted, a solution to Problem~\ref{prob:port} is immediate. The result below provides sufficient conditions for the existence of a left-inverse, thereby guaranteeing uniqueness of the solution.
\smallskip
\begin{lemma}
	Given the $Y_{jk}$ blocks in Problem~\ref{prob:port}, the matrix
	\[
	\begin{bmatrix}
		\I_{n_1} &- Y_{12} & -Y_{14} & 0 \\
		0 & -Y_{22} & -Y_{24} & 0 \\
		0& -Y_{32} & -Y_{34} & 0 \\
		0 & -Y_{42} & -Y_{44} & \I_{n_4}
	\end{bmatrix}
	\]
	has full column rank if and only if $\begin{bsmallmatrix}
		Y_{22} & Y_{24} \\ Y_{32} & Y_{34}
	\end{bsmallmatrix} $ has full column rank. 
\end{lemma}
\proof{
	Let $w \in \C^{n_1}$, $x \in \C^{n_2}$, and $y,z \in \C^{n_4}$.	Suppose that
	\begin{equation} \medmath{
		\begin{bmatrix}
			\I_{n_1} &- Y_{12} & -Y_{14} & 0 \\
			0 & -Y_{22} & -Y_{24} & 0 \\
			0& -Y_{32} & -Y_{34} & 0 \\
			0 & -Y_{42} & -Y_{44} & \I_{n_4}
		\end{bmatrix} \begin{bmatrix}
			w\\x\\y\\z
		\end{bmatrix} = \begin{bmatrix}
			w - Y_{12}x - Y_{14}y \\ -\begin{bmatrix}
				Y_{22} & Y_{24} \\ Y_{32} & Y_{34}
			\end{bmatrix} \begin{bmatrix}
				x \\ y
			\end{bmatrix}   \\ z - Y_{42}x - Y_{44}y
		\end{bmatrix} = 0.}  \label{eq:nullspace}
	\end{equation}
	If $\begin{bsmallmatrix}
		Y_{22} & Y_{24} \\ Y_{32} & Y_{34}
	\end{bsmallmatrix} $ has full column rank, then $x=y = 0$, by which $w = z = 0$. 
	If not, then there exists $ x,y$ such that $ (x,y) \neq 0$ and $\begin{bsmallmatrix}
		Y_{22} & Y_{24} \\ Y_{32} & Y_{34}
	\end{bsmallmatrix} \begin{bsmallmatrix}
		x\\ y
	\end{bsmallmatrix} = 0$. Choosing $w = Y_{12}x + Y_{14}y$ and $z = Y_{42}x + Y_{44}y$ then satisfies \eqref{eq:nullspace} for $(w,x,y,z) \neq 0$. \hfill\QED}
\smallskip
\begin{corollary}[Uniqueness of estimates] \label{cor:uniqueness}
	If $\begin{bsmallmatrix}
		Y_{22} & Y_{24} \\ Y_{32} & Y_{34}
	\end{bsmallmatrix} $ has full column rank, then Problem~\ref{prob:port} has at most a single solution. 
	If it is not full column-rank, then there are either an infinite number of solutions, or none at all. 
\end{corollary}
\smallskip
In the full column-rank case, it is still possible that no solutions exist, due to measurement and modelling errors. The Moore-Penrose pseudoinverse provides some robustness against these errors by returning the least-squares solution. It is utilised below in Definition~\ref{def:theta}, which introduce a matrix operator that maps the known signals to the fault-port signals.
\smallskip
\begin{defn}[Fault signal estimator] \label{def:theta}
	Given a matrix ${Y:= \begin{bsmallmatrix} Y_{jk} \end{bsmallmatrix}_{j,k=1}^4}$ with blocks ${Y_{jk} \in \C^{n_j \times n_k}}$, define the operator ${\Psi:\C^{N\times N} \to \C^{2n_4 \times (n_1 + n_2 + 2n_3)}}$ as
	\begin{multline} 
		\Psi(Y):=\begin{bmatrix} 0 & 0 & 0 & 0 \\ Y_{41} & 0 & Y_{43} & 0
		\end{bmatrix} \\ 
		 -\begin{bmatrix}0 & \I_{n_4} \\  Y_{42} & Y_{44} \end{bmatrix}  \begin{bmatrix}
			Y_{22} & Y_{24} \\ Y_{32} & Y_{34}
		\end{bmatrix} ^\dagger\begin{bmatrix} 	Y_{21} & -\I_{n_2}  & Y_{23} & 0 \\
			Y_{31} & 0 & Y_{33} & -\I_{n_3} 
		\end{bmatrix} \,, \label{eq:PsiY}
	\end{multline}
	where ${N:= n_1 + n_2 + n_3 + n_4}$.
\end{defn}
\smallskip
Under the rank condition given in Corollary~\ref{cor:uniqueness}, a closed-form expression for the fault-port signals is now derived, in terms of the matrix operator $\Psi(Y)$ given in~\eqref{eq:PsiY}.
%
%
\smallskip
\begin{theorem} \label{thm:estimator}
	Suppose $(i_1,v_2,v_4,i_4)$ is a solution to Problem~\ref{prob:port}. If $\begin{bsmallmatrix}
		Y_{22} & Y_{24} \\ Y_{32} & Y_{34}
	\end{bsmallmatrix} $ has full column rank, then 
	$$ \begin{bsmallmatrix} v_4 \\ i_4 \end{bsmallmatrix} = \Psi(Y)  \begin{bsmallmatrix}
		v_1 \\ i_2 \\ v_3 \\ i_3
	\end{bsmallmatrix}.$$
\end{theorem}
\proof{
	By the middle two rows of \eqref{eq:YiYv} in Lemma~\ref{lem:rearrange},
	$$ -\begin{bmatrix}
		Y_{22} & Y_{24} \\ Y_{32} & Y_{34}
	\end{bmatrix} \begin{bmatrix}
		v_2 \\ v_4
	\end{bmatrix} = \begin{bmatrix} 	Y_{21} & -\I_{n_2}  & Y_{23} & 0 \\
		Y_{31} & 0 & Y_{33} & -\I_{n_3} 
	\end{bmatrix}\begin{bsmallmatrix}
		v_1 \\ i_2 \\ v_3 \\ i_3
	\end{bsmallmatrix}.$$
	If $\begin{bsmallmatrix}
		Y_{22} & Y_{24} \\ Y_{32} & Y_{43}
	\end{bsmallmatrix} $ has full column rank, $\begin{bsmallmatrix}
		Y_{22} & Y_{24} \\ Y_{32} & Y_{43}
	\end{bsmallmatrix}^\dagger$ is a left-inverse, and therefore
	\begin{equation} \begin{bmatrix}
			v_2 \\ v_4
		\end{bmatrix} = - \begin{bmatrix}
			Y_{22} & Y_{24} \\ Y_{32} & Y_{34}
		\end{bmatrix} ^\dagger\begin{bmatrix} 	Y_{21} & -\I_{n_2}  & Y_{23} & 0 \\
			Y_{31} & 0 & Y_{33} & -\I_{n_3} 
		\end{bmatrix}\begin{bsmallmatrix}
			v_1 \\ i_2 \\ v_3 \\ i_3
		\end{bsmallmatrix}. \label{eq:v2v4} \end{equation}
	Now, by the final row of \eqref{eq:YiYv},
	\begin{align*} 
		i_4 & = \begin{bmatrix} Y_{41} & 0 & Y_{43} & 0
		\end{bmatrix} \begin{bsmallmatrix}
			v_1 \\ i_2 \\ v_3 \\ i_3
		\end{bsmallmatrix}  + \begin{bmatrix} Y_{42} & Y_{44} \end{bmatrix} \begin{bmatrix}
			v_2 \\ v_4
		\end{bmatrix},
	\end{align*}
	and therefore,
	\begin{align*}
		\begin{bmatrix} v_4 \\ i_4 \end{bmatrix}	& = \begin{bmatrix} 0 & 0 & 0 & 0 \\ Y_{41} & 0 & Y_{43} & 0
		\end{bmatrix} \begin{bsmallmatrix}
			v_1 \\ i_2 \\ v_3 \\ i_3
		\end{bsmallmatrix}  + \begin{bmatrix}0 & \I_{n_4} \\  Y_{42} & Y_{44} \end{bmatrix} \begin{bmatrix}
			v_2 \\ v_4
		\end{bmatrix}.
	\end{align*}
Substituting \eqref{eq:v2v4} then yields the result. \hfill\QED}
\smallskip

The next result provides a bound on $\| \Psi(Y) \| $.

\smallskip
\begin{lemma} \label{lem:bound}
	Letting $Y \in \C^{N \times N}$ be as in Definition~\ref{def:theta},
\[	
\| \Psi(Y) \|  \leq \| Y \| + \frac{(\|Y\|+1)^2 }{\sigma_{n_2 + n_4}(\underline{Y})} \, ,
\]
where $\underline{Y}:= \begin{bsmallmatrix}
	Y_{22} & Y_{24} \\ Y_{32} & Y_{34}
\end{bsmallmatrix}$. 
\end{lemma}
\proof{Recalling Definition~\ref{def:theta},
	\begin{align*}
		 \begin{Vsmallmatrix} 0 & 0 & 0 & 0 \\ Y_{41} & 0 & Y_{43} & 0
		\end{Vsmallmatrix} &\leq \| Y \|\\
		 \begin{Vsmallmatrix}0 & \I_{n_4} \\  Y_{42} & Y_{44} \end{Vsmallmatrix} \leq  \begin{Vsmallmatrix}0 & 0 \\  Y_{42} & Y_{44} \end{Vsmallmatrix} + \begin{Vsmallmatrix}0 & \I_{n_4} \\  Y_{42} & Y_{44} \end{Vsmallmatrix} & \leq \|Y\| + 1 \\
	\begin{Vsmallmatrix} Y_{21} & -\I_{n_2}  & Y_{23} & 0 \\
			Y_{31} & 0 & Y_{33} & -\I_{n_3} 
		\end{Vsmallmatrix} &\leq \|Y\| + 1.  \end{align*} 
	If $\underline{Y}$ is full column-rank, then $\|  \underline{Y}^\dagger \| = \frac{1}{\sigma_{n_2 + n_4}(\underline{Y})}$, where
	$\sigma_{n_2 + n_4}(\underline{Y})>0$, and the result then follows by the triangle inequality.
	Otherwise, $\sigma_{n_2 + n_4}(\underline{Y}) = 0$, and the result holds trivially. \hfill\QED}
\subsection{Signal estimation}
Applying Theorem~\ref{thm:estimator} to Clause \ref{cl:attruth} of Assumption~\ref{ass:model} yields the frequency-domain estimates
\begin{equation}  \begin{bmatrix}
	\hat{V}_4(s;d) \\ \hat{I}_4(s;d) 
\end{bmatrix} := \Psi\big(Y(s;d)\big) \begin{bsmallmatrix}
	V_1(s) \\ I_2 (s)\\ V_3(s) \\ I_3(s)
\end{bsmallmatrix}, \label{eq:est}  \end{equation}
for any $s$ at which the full column-rank condition holds. To obtain time-domain estimates, the inverse Laplace transform is now required. In order for it to be well-defined, the rate of growth of $\Psi(Y(s;d))$ must be bounded. As demonstrated by Lemma \ref{lem:bound}, this requires a lower bound on the ${\underline{Y} \in \C^{(n_2 + n_3) \times (n_2 + n_4)}}$ submatrix. Bounding its $(n_2 + n_4)$th singular value ensures it remains full column-rank. 

\begin{assumption}\label{ass:sigmabound}
Given the blocks ${Y_{jk}(s) \in \C^{n_j \times n_k}}$ that partition ${Y(s;d) = \begin{bsmallmatrix}
		Y_{jk}(s;d)
	\end{bsmallmatrix}_{j,k=1}^4 }$, the submatrix \begin{equation}
		\underline{Y}(s;d):=\begin{bsmallmatrix}
			Y_{22}(s;d) & Y_{24}(s;d) \\ Y_{32}(s;d) & Y_{34}(s;d)
		\end{bsmallmatrix}
	\end{equation}
is lower-bounded as follows: There exists $M' > 0$ such that,
\begin{equation}  \forall s \in \overline{\C^+_\beta}\ \forall d \in \D,\  \sigma_{n_2 + n_4}( \underline{Y}(s;d)) \geq M'e^{ - \tau \Re(s) }. \label{eq:sigmabound} \end{equation}
\end{assumption}
The constant $\tau$ above is from Assumption~\ref{ass:model}. There is no loss of generality here, because if \eqref{eq:normbound} and \eqref{eq:sigmabound} hold for different values of $\tau$, the largest can simply be chosen.
\smallskip
\begin{remark}[Sensing information]\label{rem:robustness}
	Assumption~\ref{ass:sigmabound} relates to the availability and positioning of sensors at Ports 2, 3 and 4 in that it guarantees the availability of a sufficient number of independent measurements to localise the fault. Since the $\underline{Y}(s,d)$ submatrix is full-column rank, this implies $n_3 \geq n_4$. Thus, the number of scalar ports at which both voltage and current are measured can be no less than the dimension of the fault port. 
\end{remark}

Under Assumption~\ref{ass:sigmabound}, an exponential bound on $\Psi(Y(s;d))$ can be derived, similar to that of $Y(s;d)$ in Assumption~\ref{ass:model}.
\begin{lemma} \label{lem:expBound}
	Under Assumptions \ref{ass:wellposed}, \ref{ass:model} and \ref{ass:sigmabound}, there exists $K \geq 5$ such that,
	$$ \forall s \in \overline{\C^+_\beta} \ \forall d \in \D,\ \| \Psi(Y(s;d)) \| \leq  K e^{ 3 \tau \Re(s)}. $$
\end{lemma}
\proof{
Choose any $s \in \overline{\C^+_\beta}$ and $d \in \D$. Assumptions~\ref{ass:model} and~\ref{ass:sigmabound} imply that there exist $M'>0$ and $M \geq 1$ such that
\begin{align*} X&:=\max\{M, \frac{1}{M'} \} e^{\tau \Re(s)} \\
& \geq \max \left\{ \| Y(s;d) \|,\ \frac{1}{\sigma_{n_2 + n_4}( \underline{Y}(s;d)) } \right\}. \end{align*}
 Applying Lemma \ref{lem:bound}, we obtain
\begin{align*} \| \Psi(Y(s;d)) \| &\leq X + (X+1)^2 X \\
	&\leq X^3 + 2X^2 + 2X \\
		&\leq X^3 + 2X^3 + 2X^3 = 5X^3,
\end{align*}
because $X \geq 1$. 
 \hfill \QED}	
\smallskip

The standard inversion result~\cite[Chapter 3]{dullerud_course_2000}, based on the Paley-Weiner theorem, considers transfer functions in $\mathcal{H}_\infty$ operating on frequency domain signals in $\mathcal{H}_2$. To accommodate unbounded growth of the transfer function in the right half-plane, some modifications are required. 
\smallskip
\begin{proposition}[Laplace inversion for acausal systems] \label{prop:inversion}
Let $n,m \in \N^+$, $a \in \R$, $t_0 \geq 0$ and $b > a$. Suppose also that ${P: \C^+_a \to \C^{m \times n}}$ is analytic, $u \in \Ex^n(a)$ and
\begin{equation}
	\forall t \in [0,t_0),\ u(t) = 0.\label{eq:utau}
\end{equation}
If there exists $K > 0$ such that
\begin{equation} \forall s \in \overline{\C^+_b},\ \| P(s) \| \leq  Ke^{t_0 \Re(s)}, \label{eq:Pbound} \end{equation}
then the inverse Laplace transform of 
	\begin{equation}
		F:\C^+_a \to \C^m,\	F(s) := P(s)U(s) \label{eq:Ydef}
	\end{equation}
	is the measurable function $f:[0,\infty) \to \R^m$ given by
	\begin{equation} f(t) = \frac{e^{ b t}}{2 \pi}\int_{- \infty}^\infty P(b +\j \omega) U(b  + \j \omega) e^{\j \omega t} \d \omega. \label{eq:IFT} \end{equation}
\end{proposition}
\proof{
	Suppose \eqref{eq:Pbound} holds, and define
	\[
	 w(t):= e^{- b t} u(t + t_0) \, ,
	 \]
	  for all $t \geq 0$. Then, $w \in \Ex^n(a - b) \subset \mathcal{L}_2[0,\infty)$, because $b > a$ and \eqref{eq:utau} holds. Therefore, $W = \Lt[w]\in \mathcal{H}_2$, where $W:\C^+_{a - b} \to \C^n$ is given by
	\begin{equation}
		W(s) = e^{t_0 (s+b)}U(s + b), \label{eq:Wdef}
	\end{equation}
	which follows from the translation properties of the Laplace transform.
	Now the transfer function $Q:\C^+_{a-b} \to \C^{m \times n}$ given by \begin{equation} Q(s):= e^{- t_0 (s+b)} P(s + b) \label{eq:Qdef} \end{equation}
		is analytic. Moreover, for all $s \in \C^+$, 
	\begin{equation} \| Q(s) \| \leq  e^{- t_0 \Re(s + b)} \|P(s + b) \| \leq K  \end{equation}
	by \eqref{eq:Pbound}, by which $Q \in \mathcal{H}^{m \times n}_\infty.$ Define $G:\C^+_{a - b} \to \C^m$ as \begin{align} G(s)&:=Q(s)W(s) \nonumber \\
		& = P(s + b)U(s+b) = F(s + b),
		\label{eq:Zdef} \end{align} 
		where $F$ is defined in \eqref{eq:Ydef}.
		Since $W \in \mathcal{H}_2$ and $Q \in \mathcal{H}_\infty$, it follows that $G \in \mathcal{H}_2$ by \cite[Theorem 2(ii), Chapter 2]{francis_course_1987}. Thus, there exists $g \in \mathcal{L}_2[0,\infty)$ such that $G = \Lt[g]$ by \cite[Theorem 3.20(b)]{dullerud_course_2000}, and it is given by the inverse Fourier transform
	\begin{equation}
		\forall t \geq 0,\	g(t) = \frac{1}{2\pi} \int_{-\infty}^\infty Q(\j \omega)W(\j \omega)e^{\j \omega t} \d \omega. \label{eq:zdef}
	\end{equation}
Applying the translation property once again, \eqref{eq:Zdef} implies the inverse Laplace transform of $F$ satisfies $f(t)= e^{b t}g(t) $ for all $t \geq 0$. Equation~\eqref{eq:IFT} then follows from \eqref{eq:zdef}, \eqref{eq:Wdef} and \eqref{eq:Qdef}. Since $g \in \mathcal{L}_2[0,\infty)$, $f$ is also measurable. \hfill \QED}

\smallskip
The difference between Proposition~\ref{prop:inversion} and the standard Paley-Weiner is \eqref{eq:utau}, which constrains the input to be zero before time $t_0 \geq 0$. No such constraint applies to the output, potentially making transfer function $P$ acausal.
\smallskip
\begin{remark}[Bound on lead-time] \label{rem:delay}
	The space $\mathcal{H}_\infty$ corresponds to the set of stable and causal linear time-invariant systems~\cite[Chapter 3.4.3]{dullerud_course_2000}, and the transfer functions therein are bounded over $\C^+$. If a transfer function is analytic over $\overline{\C}^+$, but grows exponentially as $\Re(s) \to \infty$, it is stable and time-invariant, but not causal. Proposition~\ref{prop:inversion} demonstrates that the rate of exponential growth ``quantifies" the degree of acausality by bounding the time by which the system output can precede the input. This is why $\tau$, in Assumption~\ref{ass:model}, upper-bounds the propagation delay from the fault ports to the known ports.
\end{remark}
\smallskip
\begin{remark}[Numerical evaluation]
	The inversion in \eqref{eq:IFT} can be performed numerically by inverse FFT. This involves the evaluation of $U(s)$ along the vertical line $\Re(s)=b$ in the complex plane, which in turn amounts to pointwise multiplication of the original signal $u$ by the exponential window $e^{-bt}$, followed by an FFT. Choosing $b>0$ makes it a decaying exponential, which improves numerical accuracy because it attenuates the effect of having a finite measurement horizon. 
\end{remark}
\smallskip
To apply Proposition \ref{prop:inversion} to \eqref{eq:est}, the condition \eqref{eq:utau} must be met for all the known port signals. It can be assumed without loss of generality, because it only requires a shift of the time origin to the left. 

\begin{assumption} \label{ass:zerooo}
	The known signal $u:[0,\infty) \to \R^{n_1 + n_2 + 2n_3}$ defined as
	$ u(t): = \left(	v_1(t), i_2(t), v_3(t), i_3(t) \right) $
	satisfies $ u(t) = 0$ for all $t \in [0, 3\tau)$.
\end{assumption}
\smallskip

The main result of the paper can  now be stated. It constructs fault-port signal estimates in both the time and frequency domains, for every candidate fault location $d$. Moreover, it guarantees the estimates are equal to the true signals when evaluated at the true fault location $\ell$. Its proof formalises the argument of the paper up to this point. 
\smallskip
\begin{theorem}[Fault signal estimates] \label{thm:signalEst}
	Under Assumptions~\ref{ass:wellposed}, \ref{ass:model}, \ref{ass:faultModel}, \ref{ass:sigmabound}, and \ref{ass:zerooo},
	define the estimates $\hat{V}_4, \hat{I}_4:\C^+_\alpha \times \D \to \C^{n_4}$ and ${\hat{V}_f:\C^+ \times \D \to \C^{n_4}}$ as follows:
	\begin{subequations}
		\begin{align}
			\begin{bmatrix}
				\hat{V}_4(s;d) \\ \hat{I}_4(s;d) 
			\end{bmatrix} &:= \Psi\big(Y(s;d)\big) U(s)\label{eq:IVest}, \\
			\hat{V}_f(s;d) &:= - \Phi(s) \hat{I}_4(s;d). \label{eq:Vfest}
		\end{align} 
		\label{eq:freqest}
	\end{subequations}
	The corresponding inverse Laplace transforms ${\hat{v}_4,\hat{\i}_4,\hat{v}_f:[0,\infty) \times \D \to \R^{n_4}}$, given by 
	\begin{equation}
		\begin{bsmallmatrix}
			\hat{v}_4(t;d) \\ \hat{\i}_4(t;d) \\ \hat{v}_f(t;d)
		\end{bsmallmatrix} = \frac{e^{ \beta t}}{2 \pi}\int_{- \infty}^\infty\begin{bmatrix}
			\hat{V}_4(\beta + \j \omega;d) \\ \hat{I}_4(\beta + \j \omega;d)  \\ \hat{V}_f(\beta + \j \omega;d)
		\end{bmatrix}e^{\j \omega t} \d \omega, \label{eq:timeEst}
	\end{equation}
	are measurable in $t$ for every $d$. In the frequency domain,
	\begin{subequations}
		\begin{align}
			\forall s \in \C^+_\alpha,\quad	&\hat{V}_4(s;\ell) = V_4(s),\ \hat{I}_4(s;\ell) = I_4(s), \label{eq:AtTruthFreq} \\
			\forall s \in \C^+,\	& \hat{V}_f(s;\ell) = V_f(s),  \label{eq:AtTruthFreqF}
		\end{align} \label{eq:AtTruth}
	\end{subequations}
and in the time domain,
	\begin{align}
		&\hat{v}_4(t;\ell) = v_4(t), && \hat{\i}_4(t;\ell) = i_4(t), && \hat{v}_f(t;\ell) = v_f(t),\label{eq:AtTruthTime}
	\end{align}
	for almost every $t \geq 0$.
\end{theorem}
\proof{
	For every $k \in \{1,...,4\}$, the signals $v_k,i_k \in \Ex(\alpha)$ by Assumption~\ref{ass:wellposed}, so their Laplace transforms $V_k,I_k$ converge on $\C^+_\alpha$. Assumption~\ref{ass:sigmabound} implies $ \underline{Y}(s;\ell) $ is full column rank for all $s \in \overline{\C^+_\beta}$. Since $I(s) = Y(s;\ell) V(s)$  by Assumption~\ref{ass:model}, \eqref{eq:IVest} implies \eqref{eq:AtTruthFreq} by Theorem~\ref{thm:estimator}. Equation \eqref{eq:AtTruthFreqF} then follows from \eqref{eq:Vfest}, because $V_f(s) = - \Phi(s) I_4(s)$ by Assumption~\ref{ass:faultModel}. Equation \eqref{eq:freqest} also implies $$ \hat{V}_f(s;d) = \overbrace{\begin{bmatrix}
		0 & -\Phi(s)
	\end{bmatrix} \Psi(Y(s;d)) }^{\Psi_f(s;d)} U(s) ,$$
	making $\Psi_f(s;d)$ the transfer function from the known signals to the estimate $\hat{V}_f$. 
	
	For every $d$, $Y(s;d)$ is analytic in $s$, and the same holds for $\Psi(Y(s;d))$, because the matrix operations in Definition~\ref{def:theta} preserve analyticity. By Lemma \ref{lem:expBound}, for all $s \in \overline{\C^+_\beta}$ and $d \in \D$,
	$$ \| \Psi(Y(s;d)) \| \leq K e^{3 \tau \Re(s)}.$$
Since $\Phi \in \mathcal{H}_\infty$, $$\| \Psi_f(s;d) \| \leq \| \Phi\|_\infty K e^{3 \tau \Re(s)}$$ 
	for all $d \in \D$ and $s \in \overline{\C^+_\beta}$, as well. Recalling Assumption \ref{ass:zerooo}, all requirements of Proposition~\ref{prop:inversion} have thus been met. The signals $\hat{v}_4,\hat{\i}_4, \hat{v}_f$ in \eqref{eq:timeEst} are therefore inverse Laplace transforms of $\hat{V}_4, \hat{I}_4, \hat{V}_f$, and are measurable in $t$ for every $d \in \D$. The Laplace transform has a unique inverse modulo sets of measure zero~\cite[Theorem 5.1]{doetsch_introduction_1974}, so \eqref{eq:AtTruthTime} follows from both \eqref{eq:AtTruth} and Assumption~\ref{ass:wellposed}. \hfill\QED}
\smallskip

It is worth noting that Assumption~\ref{ass:continuous} has not been invoked thus far, because even when $Y(s;d)$ is continuous in $d$, the same cannot be guaranteed for the signal estimates.  Such continuity simplifies numerical implementation of the subsequent localisation step, and is assured under the following additional assumption on the known port signals.
\smallskip
\begin{assumption}[Smoothness] \label{ass:smoothness}
	The known signal $u$ satisfies
	\begin{align*} 
		\int_{- \infty}^\infty \| U(\beta + \j \omega) \|_1 \d \omega < \infty.
	\end{align*}
\end{assumption}
\smallskip
\begin{remark}[Continuity] \label{rem:continuity}
	Assumption~\ref{ass:smoothness} requires the Fourier transforms of the known signals $v_1, i_2, v_3, i_3$ to be in $\mathcal{L}_1(\j\R)$, after pointwise multiplication by $e^{-\beta t}$ in the time-domain. This relates to their smoothness~\cite[Section 8.4]{folland_real_1999}.
	Assumption~\ref{ass:wellposed} alone does not ensure this, although various other sufficient conditions have been derived. For example, Assumption~\ref{ass:smoothness} holds if the signals are bandlimited, or belong to the Schwartz space. For more comprehensive treatments, see~\cite{herz_lipschitz_1968,moricz_sufficient_2010,iosevich_decay_2014}. Since the known port signals are affected by the fault (indeed, localisation relies on this), it is difficult to know whether such conditions hold in a given practical case. If not, the cost functions constructed in the next section may not be continuous, making their minimisation more challenging. 
\end{remark}
\smallskip
\begin{lemma}[Continuous estimates] 
	\label{lem:continuity}
	If the hypotheses of Theorem~\ref{thm:signalEst} hold, along with Assumptions~\ref{ass:continuous} and~\ref{ass:smoothness}, then $\hat{v}_4(t;d), \hat{\i}_4(t;d)$ and $\hat{v}_f(t;d)$ are continuous in $d$ for every $t \geq 0$.  
\end{lemma}
\proof{
	The estimates $\hat{V}_4(s;d), \hat{I}_4(s;d)$ and $\hat{V}_f(s;d) $ are continuous in $d$, because $Y(s;d)$ is continuous in $d$ by Assumption~\ref{ass:continuous}, and $\Psi$ is continuous by Definition~\ref{def:theta}. By Lemma~\ref{lem:expBound}, $$ \forall d \in \D,\ \forall \omega \in \R,\ \| \Psi (Y(\beta + \j \omega;d)) \| \leq K e^{3\tau \beta},$$ and so $\| \Psi (Y(\beta + \j \omega;d)) \|_{1,1}$ is also bounded over $\omega \in \R$ and $d \in \D$.  Similarly, $\| \Phi(\beta + \j \omega)\|$ is bounded over $\omega \in \R$ by Assumption~\ref{ass:faultModel}.  Thus, by \eqref{eq:freqest}, there exists $C > 0 $ such that $\forall \omega \in \R,\ \forall d \in \D,$
	$$ \begin{Vsmallmatrix}
		\hat{V}_4(\beta + \j \omega;d) \\ \hat{I}_4(\beta + \j \omega;d) \\ \hat{V}_f(\beta + \j \omega;d)
	\end{Vsmallmatrix}_1 \leq C  \| U(\beta + \j \omega)\|_1.$$
	Continuity of $\hat{v}_4(t;d), \hat{\i}_4(t;d)$ with respect to $d$ then follows from both \eqref{eq:timeEst} and Assumption~\ref{ass:smoothness}, by the Dominated Convergence Theorem~\cite[Theorem 3.31]{axler_measure_2020}. \hfill\QED}
 
\subsection{Optimisation}
Given Theorem~\ref{thm:signalEst}, Problem~\ref{prob:knownPhi} is easily reformulated as an optimisation problem. Below, a cost is assigned to every candidate fault location $d$ and candidate occurrence time $\theta$, recalling that the true fault time $t_f$ must satisfy Assumption~\ref{ass:faultModel}. 
\smallskip
\begin{corollary}[Known fault impedance] \label{cor:knownPhi}
	Let $T > T_f$ and define the cost function ${J:\D \times [0,T_f] \to [0,\infty)}$ as
	\begin{equation*}
		\medmath{ J(d,\theta):= \int_0^\theta \| \hat{v}_f(t;d) \|^2 \d t + \int_\theta^T \|\hat{v}_f(t;d) -  \hat{v}_4(t;d)  \|^2 \d t.}
	\end{equation*}
Under the hypotheses of Theorem~\ref{thm:signalEst}, we have $J(\ell,t_f) = 0$.
	Moreover, if Assumptions~\ref{ass:continuous} and~\ref{ass:smoothness} hold, then $J$ is continuous. 
\end{corollary}
\proof{
	The fact that $J(\ell,t_f) = 0$ is a direct consequence of~\eqref{eq:AtTruthTime} and \eqref{eq:switch}. The continuity of $J$ with respect to $d$ follows from Lemma~\ref{lem:continuity} by the Bounded Convergence Theorem~\cite[Theorem 3.26]{axler_measure_2020}. Continuity with respect to $\theta$ is implied by \cite[Theorem 11, Chapter 6]{royden_real_2010}. \hfill\QED}
\smallskip

The cost function $J$ must now be minimised with respect to both distance and occurrence time. Corollary~\ref{cor:knownPhi} guarantees the existence of a global minimum at the true values, even in the absence of continuity. 
Thus,
\begin{equation} 
(\ell, t_f) \in  \argmin_{d,\theta} J(d,\theta) \, , \label{eq:Jsol} 
\end{equation}
that is, this set of candidate solutions to Problem~\ref{prob:knownPhi} is guaranteed to contain the true values.

Clearly, the above approach requires the knowledge of the impedance $\Phi$ to construct the fault voltage estimate $\hat{V}_f$ in \eqref{eq:Vfest}. In the absence of a fault impedance model, an alternative fault voltage estimate $\tilde{V}_f$ can be recovered from~\eqref{eq:switch}, and Assumption~\ref{ass:passive} then exploited by means of a different cost function. Its definition uses the non-negative projection operator $(\cdot)^+$, defined in \eqref{eq:xplus}.
\smallskip
\begin{theorem}[Passive fault impedance] \label{thm:passivePhi}
	Suppose that Assumption~\ref{ass:passive} holds, along with the hypotheses of Theorem~\ref{thm:signalEst}. Define $\tilde{V}_f:\C^+_\alpha \times \D \times [0,T_f]$ as
	\begin{equation}  \tilde{V}_f(s;d, \theta) := \int_\theta^\infty  \hat{v}_4(t;d) e^{- s t} \d t. \label{eq:Uhat} \end{equation}
	Let $\omega_B> 0$, and define ${\tilde{J}:\D \times [0,T_f] \to [0,\infty)}$ as
	\begin{equation} \tilde{J}(d,\theta) := \int_{- \omega_B}^{\omega_B} \Re[\hat{I}_4(\beta + \j \omega; d )^* \tilde{V}_f(\beta + \j \omega; d,\theta)]^+ \d \omega. \label{eq:Jtil} \end{equation}
	Then $\tilde{J}(\ell,t_f) = 0$. Moreover, if Assumptions~\ref{ass:continuous} and~\ref{ass:smoothness} hold, then $\tilde{J}$ is continuous. 
\end{theorem} 
\proof{
	Since $\Phi$ is positive real by Assumption~\ref{ass:passive}, we have
	\begin{align*} 
		\Re [I_f(s)^* V_f(s)] & = \Re[I_f(s)^* \Phi(s) I_f(s) ] \\
		& =  I_f(s)^* \left( \frac{\Phi(s)^* + \Phi(s)}{2} \right) I_f(s) \geq 0 \, ,
	\end{align*}
	$\forall s \in \C^+ \cap \C_\alpha$.
	In particular, recalling that $i_4 = - i_f$, we get
	\[
	 \forall \omega \in \R,\ \Re[I_4(\beta + \j \omega)^* V_f(\beta + \j \omega)] \leq 0 \, .
	 \]
	Since $ I_4(s) = \hat{I}_4(s;\ell) $ by \eqref{eq:AtTruthFreq},  and $V_f(s) = \Lt[v_f](s) = \tilde{V}_f(s;\ell,t_f)$ by \eqref{eq:switch}, it follows that
\[
 \forall \omega \in \R,\ \Re[\hat{I}_4(\beta + \j \omega; \ell )^* \tilde{V}_f(\beta + \j \omega; \ell,t_f)]^+ = 0 \, .
 \]
Hence $\tilde{J}(\ell,t_f) = 0$. To establish continuity of $\tilde{J}$, 
	observe that
	\begin{align}  \hat{V}_4(s;d) &= \int_0^\infty \hat{v}_4(t;d)e^{- s t} \d t \nonumber
		\\&= \tilde{V}_f(s;d,\theta) + \int_0^\theta \hat{v}_4(t;d) e^{-st} \d t, \label{eq:splitIntegral}\end{align}
	by Theorem~\ref{thm:signalEst}. By \eqref{eq:freqest}, $\hat{I}_4$ and $\hat{V}_4$ are continuous under Assumption~\ref{ass:continuous}. Continuity of $\tilde{V}_f$ then follows from that of $\hat{v}_f$ in $d$ by Lemma~\ref{lem:continuity}, and continuity of $\tilde{J}$ from the Bounded Convergence Theorem ~\cite[Theorem 3.26]{axler_measure_2020} and \cite[Theorem 11, Chapter 6]{royden_real_2010}. \hfill\QED}
\smallskip

Just as for Problem~\ref{prob:knownPhi}, the cost $\tilde{J}$ must be minimised with respect to both distance and occurrence time to yield a set of candidate solutions to Problem~\ref{prob:passivePhi}. If the fault impedance is known, both techniques can be leveraged to narrow down the set of solutions:
\begin{equation} (\ell, t_f) \in  \argmin_{d,\theta} J(d,\theta) \cap  \argmin_{d,\theta}  \tilde{J}(d,\theta). \label{eq:tight_sol} \end{equation}
\smallskip
\subsubsection{Estimation of fault occurrence time}
Minimisation of $J$ or $\tilde{J}$ is a non-convex optimisation problem over a 2-dimensional box. This can be reduced to a single dimension by first obtaining the fault time as a function of the distance. A fault time estimate $\hat{\theta}(d)$ can be easily obtained from the current estimate $\hat{\i}_4(t;d)$, by finding its activation time (i.e., the time at which it first becomes non-zero over some open interval).
\smallskip
\begin{lemma} \label{lem:faultTime}
	Under the hypotheses of Theorem~\ref{thm:signalEst}, define
	${\hat{\theta}:\D \to [0,\infty]}$ as
	\begin{equation} \hat{\theta}(d):= \inf\left\{ \theta \geq 0 \mid \int_0^\theta |\hat{\i}_4(t;d)| \d t > 0\right\}. \label{eq:Tf} \end{equation}
	Then $\hat{\theta}(\ell) \geq t_f$, and for almost every $t \geq 0$, we have
	\begin{equation}
		v_f(t) = \1(t - \hat{\theta}(\ell) ) v_4(t).\label{eq:switchest}
	\end{equation}
\end{lemma}
\proof{
	By \eqref{eq:Tf}, $\hat{\i}_4(t;\ell) = 0$ for almost every $t \in [0,\hat{\theta}(\ell)]$, and therefore $i_4(t) = 0$ for almost every $t \in [0,\hat{\theta}(\ell)]$ by \eqref{eq:AtTruthTime}. Now $\Phi \in \mathcal{H}^{n_4 \times n_4}_\infty$ by Assumption~\ref{ass:faultModel}, so it is causal, and since $V_f (s)= - \Phi(s) I_4(s) $ by \eqref{eq:Ohms}, this implies 
	\begin{equation}
	\text{for almost every} \, \, t \in [0,\hat{\theta}(\ell)],\ v_f(t) = 0 \, . 
	\label{eq:vf0} 
	\end{equation} 
	Choose any $t > \hat{\theta}(\ell)$. Then by \eqref{eq:Tf}, $\{ \theta \in [0,t] \mid |\hat{\i}_4(\theta;\ell)| \neq 0 \} $ has non-zero measure, and therefore $\{ \theta \in [0,t] \mid |i_4(\theta)| \neq 0 \} $ does as well. It then follows from~\eqref{eq:zerocurrent} that $t > t_f$, and thus $v_f(t) = v_4(t)$ by \eqref{eq:switch}. Together with \eqref{eq:vf0}, this establishes~\eqref{eq:switchest}. It has also been shown that $t > \hat{\theta}(\ell) \implies t > t_f$, by which $\hat{\theta}(\ell) \geq t_f$. \hfill \QED}
\smallskip

Comparing \eqref{eq:switchest} with \eqref{eq:switch} reveals that, at the true fault location, $\hat{\theta}(\ell)$ serves as an effective fault time. It is an upper-bound on $t_f$, but need not be equal to it, because the assumptions permit the fault switch to close while the network $\cN$ is still at rest, before any of the port signals have activated. If this is the case, then $\hat{\theta}(\ell)$ is the time at which Port 4 activates, rather than when the switch closes.
\smallskip
\begin{corollary}
	Suppose the hypotheses of Theorem~\ref{thm:signalEst} hold. \begin{itemize} 
		\item If $T_f \geq \hat{\theta}(\ell)$, then $J(\ell, \theta) = 0$ for all $\theta \in [t_f,\hat{\theta}(\ell)]$.
		\item If Assumption~\ref{ass:passive} holds, then $\tilde{J}(\ell,\theta)=0$ for any  ${\theta \in [t_f,\hat{\theta}(\ell)]}$.
	\end{itemize}
\end{corollary}
\proof{
	For almost every $\theta \in [t_f,\hat{\theta}(\ell)]$, $v_f(\theta)=0$ by \eqref{eq:switchest},  and
	$ v_4(\theta) = v_f(\theta) =0$ by \eqref{eq:switch}. Recalling \eqref{eq:AtTruthTime},
	$$  \hat{v}_4(\theta;\ell) = \hat{v}_f(\theta;\ell) = 0$$
	for almost every $\theta \in [t_f,\hat{\theta}(\ell)]$. 
	Thus, $J(\ell,\theta)$ is constant over this interval. The same holds for $\tilde{V}_f(s;\ell,\theta)$ in \eqref{eq:Uhat}, and consequently $\tilde{J}(\ell,\theta)$ as well. \hfill \QED}
\smallskip

Now, using the fault time estimator, the set of solutions~\eqref{eq:tight_sol} becomes
\begin{equation*} 
\ell \in  \argmin_{d \in \D} J(d,\hat{\theta}(d))  \cap  \argmin_{d \in \D}  \tilde{J}(d,\hat{\theta}(d)). \end{equation*}
%
%
\begin{remark}[Cost of dimension reduction] Continuity of  $\hat{\theta}$ is not guaranteed. If Assumptions~\ref{ass:continuous} and~\ref{ass:smoothness} hold, then minimisation of the continuous $J$ and $\tilde{J}$ over two dimensions can be replaced by minimisation of the possibly discontinuous $J(d,\hat{\theta}(d))$ and $\tilde{J}(d,\hat{\theta}(d))$ over one dimension only. In the absence of these assumptions, there may be no advantage to solving the higher-dimensional problem.
\end{remark}
\smallskip

Algorithm~\ref{alg:main} summarises the steps involved in evaluating $J(d,\theta)$, $ J(d,\hat{\theta}(d))$, $\tilde{J}(d,\theta)$ or $\tilde{J}(d,\hat{\theta}(d))$, based on the arguments provided to \textsc{LocCost}. For clarity, the algorithm is expressed in terms of exact continuous operations that are straightforward to approximate on a digital computer.
\smallskip
\begin{algorithm}[h] 
	\caption{Evaluates $J$ or $\tilde{J}$ based on information provided.} \label{alg:main}
	\begin{algorithmic}[1]	
		\Function{LocCost}{$d,\theta,T,\omega_B,\beta, v_1,i_2,v_3,i_3,Y,\Phi$} 
		\ForAll{$t \in [0,\infty)$} 
		\State $ u(t) := e^{- \beta t} \Call{VertCat}{v_1(t), i_2(t), v_3(t), i_3(t)}$ \EndFor
		\State $U := \Call{FourierTransform}{u} $ \label{ln:FTu}
		\ForAll{$\omega \in \R$}
		\State  $\begin{bsmallmatrix}
			\hat{V}_4(\omega) \\ \hat{I}_4(\omega) 
		\end{bsmallmatrix} := \Psi\big(Y(\beta + \j \omega;d)\big)U(\omega) $  \EndFor
		\State $\begin{bsmallmatrix}
			\hat{v}_4 \\ \hat{\i}_4 
		\end{bsmallmatrix} :=\Call{InverseFourierTransform}{}\left( \begin{bsmallmatrix}
			\hat{V}_4 \\ \hat{I}_4
		\end{bsmallmatrix}\right)  $ \label{ln:IFTu}
		\If{$\theta$ is not provided}
		\ForAll{$t \in [0,\infty)$}
		\State $ \hat{\i}_4(t) \leftarrow e^{\beta t} \hat{\i}_4(t)  $
		\EndFor
		\State $\theta:= \Call{ActivationTime}{\hat{\i}_4}$
		\EndIf
		\If{$\Phi$ is provided}
		\ForAll{$\omega \in \R$}
		\State  $\hat{V}_f(\omega) := -\Phi(\beta + \j \omega) \hat{I}_4(\omega)$ \EndFor
		\State $\hat{v}_f := \Call{InverseFourierTransform}{\hat{V}_f} $ \label{ln:IFTv}
		\State $\medmath{J:=\int_0^\theta e^{2 \beta t} \| \hat{v}_f(t) \|^2 \d t + \int_\theta^T e^{2 \beta t} \|\hat{v}_f(t) -  \hat{v}_4(t)  \|^2 \d t}$
		\State \Return $J$
		\Else
		\ForAll{$t \in [0,\theta)$ }
		\State $\hat{v}_4(t) \leftarrow 0 $ \EndFor
		\State $\tilde{V}_f :=\Call{FourierTransform}{\hat{v}_4}$ \label{ln:FTv}
		\State \Return $\int_{- \omega_B}^{\omega_B}  \Re[\hat{I}_4(\omega )^* \tilde{V}_f (\omega)]^+ \d \omega$ 
		\EndIf
		\EndFunction
	\end{algorithmic}
\end{algorithm}

\section{Numerical Validation} \label{sec:sim}
\subsection{Assumption verification}
The performance of Algorithm \ref{alg:main} is now evaluated numerically on Case Study \ref{ex:tline}. But it first remains to verify that the admittance matrix $Y$ constructed in Section \ref{sec:composition} satisfies Assumption~\ref{ass:sigmabound}. In Case Study \ref{ex:tline}, $n_2 + n_4 = n+1$, and the submatrix $ \underline{Y}(s;d) = \begin{bsmallmatrix}
	1 & 0 \\ 0 & -B_d(s)^{-1}
\end{bsmallmatrix}$ is square. Thus,
\begin{align*}  
	\sigma_{n_2 + n_4} (\underline{Y}(s;d) )^{-1} & = \| \underline{Y}(s;d)^{-1}   \|  = \begin{Vsmallmatrix} 1 & 0 \\ 0 & B_d(s) \end{Vsmallmatrix} \\
	& \leq \max\{ 1, \|B_d(s) \| \} \\
	& \leq M_2 e^{\frac{d}{\nu} \Re(s)},
\end{align*}
by \eqref{eq:Me2}, because $B_d$ is a block of $\Xi$. Comparing this with \eqref{eq:Ybound}, it is clear that Assumptions \ref{ass:model} and \ref{ass:sigmabound} both hold for $\tau = \frac{Ln}{\nu}$.

\subsection{Simulation setup}
\begin{figure}[t]
	\begin{subfigure}{\columnwidth}
		\includegraphics[width = \columnwidth]{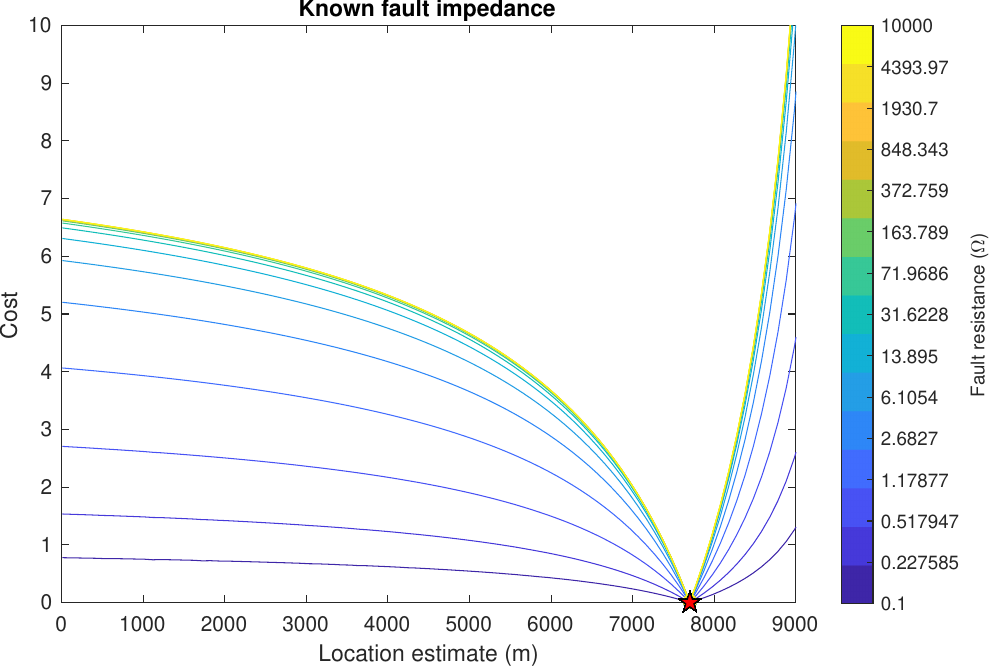}
		\caption{Plot of $J(d, \hat{\theta}(d))$.}
		\label{fig:Jknown}
	\end{subfigure}
	\begin{subfigure}{\columnwidth}
		\includegraphics[width = \columnwidth]{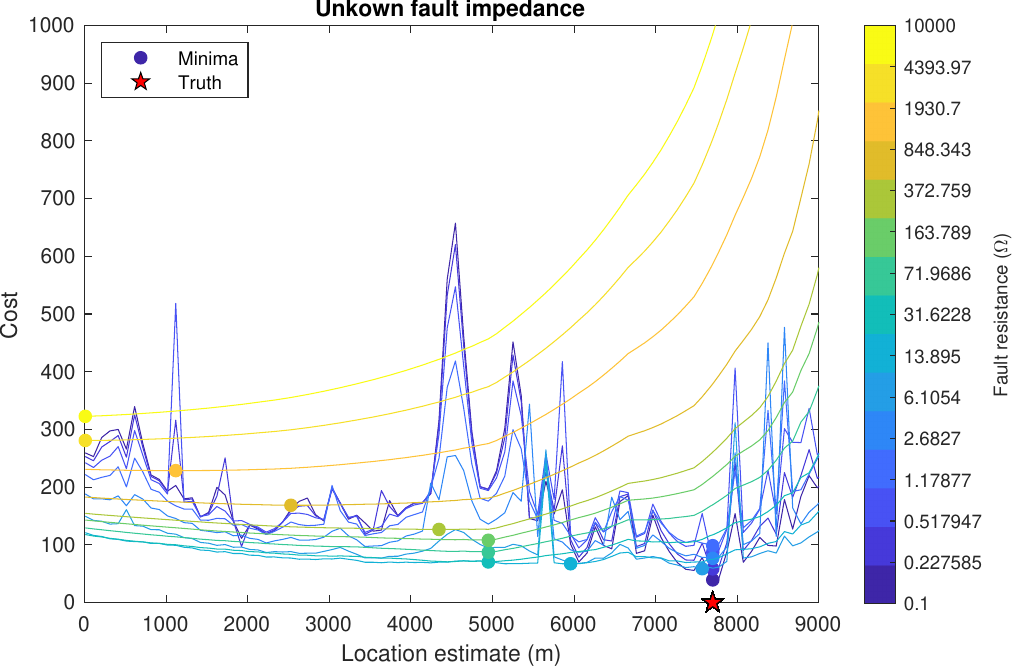}
		\caption{Plot of $\tilde{J}(d, \hat{\theta}(d))$.}
		\label{fig:Jpass}
	\end{subfigure}
	\caption{Variation of cost functions with fault resistance:  fixed fault location $\ell = 7.7$ km, and $R_f$ logarithmically spaced  from $10^{-1} \ \Omega$ to $10^4 \ \Omega$. }
	\label{fig:costfuns}
\end{figure}
PLECS~\cite{allmeling_plecs_1999} was used to simulate the transmission line. (Specifically, PLECS Blockset 4.8.1 integrated with Simulink R2023b Update 6.) Its 3-phase transmission line element~\cite{allmeling_plecs_2024} admits parameter matrices of the form
\begin{align*}
		&\Rb  = R \I_3,
	&&\Lb = \begin{bmatrix}
		L_S & L_M & L_M \\
			L_M & L_S & L_M \\
				L_M & L_M & L_S
	\end{bmatrix} ,\\
&\Gb = 0, 
&&\Cb  = \begin{bmatrix} 
	 C_E + 2 C_K & - C_K & -C_K \\
	 - C_K & C_E + 2 C_K& -C_K  \\
	 - C_K & - C_K & C_E + 2 C_K
\end{bmatrix}.
\end{align*}
The fault was triggered on only the first phase, with a constant impedance of
\begin{equation}
	\Phi(s) = R_f \I_3.
\end{equation}
A purely resistive load $R_3>0$ was also imposed at Port 3, so that
$$ v_3(t) = R_3 i_3(t),$$
and Port 1 was connected to the sinusoidal voltage source
\begin{equation}
	v_1(t) = V \begin{bmatrix}
		\sin( 2 \pi f_1 t)  \\ \sin( 2 \pi f_1 t + \tfrac{2\pi}{3}) \\ \sin( 2 \pi f_1t + \tfrac{4\pi}{3})
	\end{bmatrix}.
\end{equation}
To investigate its effect on localisation performance, the fault resistance $R_f$ was varied between $10^{-1}\ \Omega$ and $10^4\ \Omega$ at 15 logarithmically spaced intervals. Estimation performance is expected to degrade with increasing $R_f$, because Port 4 then approaches its unfaulted condition, which is an open circuit. Other parameter values are provided in Table~\ref{tab:data}. 
\begin{table}[h]
	\caption{Parameter values for Case Study \ref{ex:tline} \label{tab:data}}
	\centering
	\begin{tabular}{ |c|c|c|c| } 
		\hline
		Parameter & Symbol & Value & Units \\
		\hline 
		Simulation timestep &  & 1 & $\mu \mathrm{s}$ \\ 
		Measurement duration & $T$ & $ 4.1 $ & s \\ 
		Sensor sampling interval& & 2 & $\mu \mathrm{s}$  \\ 
		Line resistance & $R$ &  0.32 & m$\Omega/\m$ \\ 
		Line self-inductance & $L_S$ & $299.85$ & $\mathrm{nH}/\m$ \\ 
		Line mutual-inductance & $L_M$ & $59.97$ & $\mathrm{nH}/\m$ \\ 
		Line capacitances & $C_K$ & 0.33 & $\mathrm{nF}/\m$ \\ 
										& $C_E$ & 0.033 & $\mathrm{nF}/\m$ \\ 
		Line length & $L$ & 10 & km \\
		Source amplitude & $V$ & $5\sqrt{2}$ & kV \\
		Source frequency & $f_1$ & 50 & Hz \\
		Load resistance & $R_3 $ & 1 & k$\Omega$ \\
		Fault time & $t_f$ & 2.1 & s \\ 
		Fault location & $\ell$ & 7.7 & km \\ 
		Minimum fault distance & $\delta$ & 10 & m \\
		Line of complex integration & $\beta$ & 0.01 & 1 \\
		\hline
	\end{tabular}
\end{table}

Given these values, the lower-bound on propagation speed appearing in \eqref{eq:Me1}--\eqref{eq:Me2} is $\nu = 5.2 \times 10^4$~m/s  (see \cite[(21b) and Lemma 4]{selvaratnam_frequency-domain_2025}), which yields a delay bound of $\tau = 0.5337$~s. 
\subsection{Results}
 In Figure \ref{fig:costfuns}, for each value of $R_f$, the cost functions $J$ and $\tilde{J}$ are plotted as functions of distance, using the estimator proposed in \eqref{eq:Tf} to infer the fault occurrence time. In this connection, the search domain $\D = [\delta, L - \delta]$ has been discretized into 100 equispaced points, and the true location $\ell$ added in.
 The cost function $J$, which relies on accurate knowledge of the fault impedance, is plotted in Figure~\ref{fig:Jknown}. It appears continuous and quasiconvex, with a unique global minimum consistently at $\ell$. Thus, contrary to expectation, estimator performance does not degrade with $R_f$. Observe, moreover, that $J$ tends monotonically to a limiting profile as $R_f \to \infty$. 
 
\begin{figure}[t]
 	\begin{subfigure}{\columnwidth}
 		\includegraphics[width = \columnwidth]{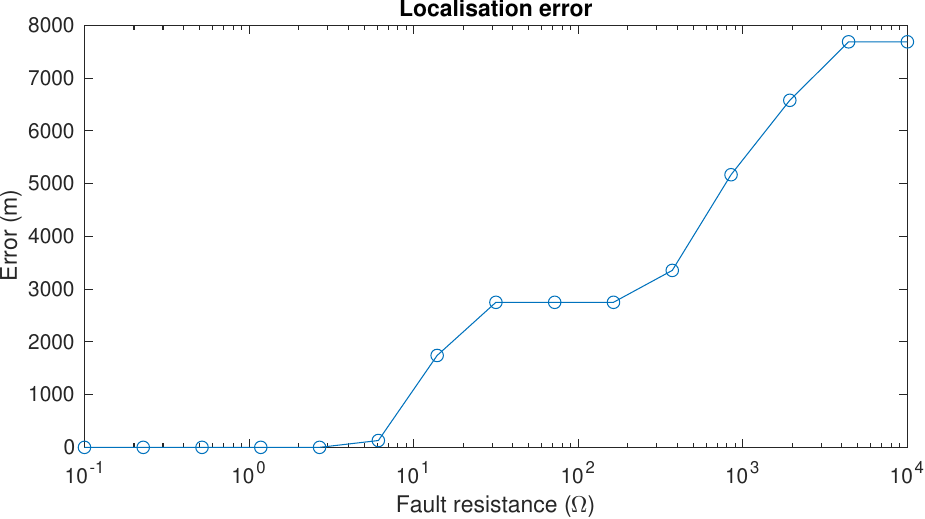}
 		\caption{Variation of $|\ell - d^\star|$ with $R_f$.}
 		\label{fig:error}
 	\end{subfigure}
 	\begin{subfigure}{\columnwidth}
 		\includegraphics[width = \columnwidth]{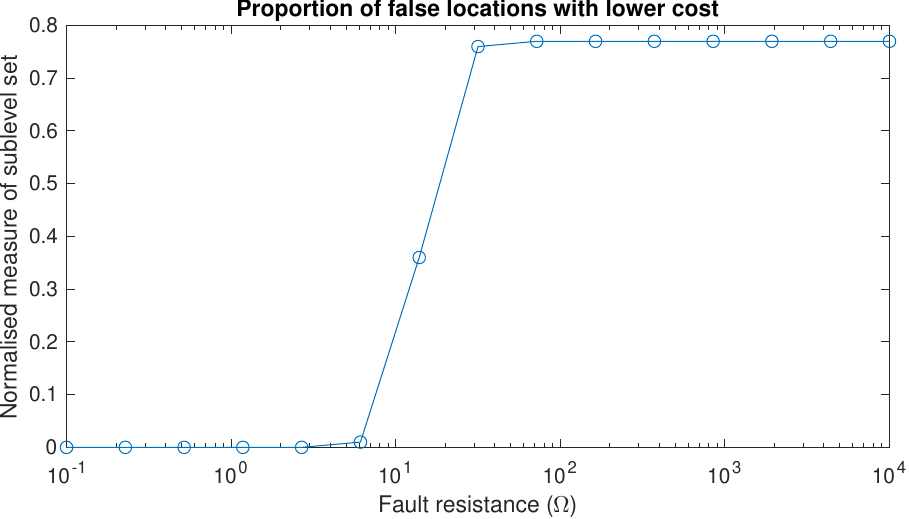}
 		\caption{Size of sublevel set $ \tilde{J}(d, \hat{\theta}(d)) \leq \tilde{J}(\ell,t_f)$.}
 		\label{fig:subop}
 	\end{subfigure}
 	\caption{Effect of unknown fault resistance on localisation performance.}
 	\label{fig:degredation}
 \end{figure}

 Cost function $\tilde{J}$, plotted in Figure \ref{fig:Jpass}, does not require any knowledge of the fault impedance other than its passivity. For low values of $R_f$, the cost function appears highly non-smooth and non-convex, but a distinct global minimum still appears at the true fault location. Thus, estimation quality is good, despite the lack of smoothness. As $R_f$ increases, $\tilde{J}$ smoothens out, but its global minimiser moves away from the truth, so estimation performance degrades. Figure \ref{fig:degredation} explicitly quantifies this degradation. In particular, the error of estimate
$$ d^\star \in \argmin_{d \in \D} \tilde{J}(d, \hat{\theta}(d) ) $$
is plotted in Figure \ref{fig:error}. Note that, in every case simulated, both $J$ and $\tilde{J}$ had unique global minimisers. Figure \ref{fig:subop} plots the proportion of the search domain with lower cost than the truth, which is given by
$$ \frac{\mathrm{Leb} \{ d \in \D \mid \tilde{J}(d, \hat{\theta}(d)) \leq \tilde{J}(\ell,t_f) \}}{\mathrm{Leb} \D },$$
where Leb denotes the Lebesgue measure. Since no measurement noise was injected, the loss of accuracy beyond $R_f \approx 5\ \Omega$ can be attributed to sampling and windowing effects when numerically approximating Fourier transforms in Lines \ref{ln:FTu}, \ref{ln:IFTu}, \ref{ln:IFTv} and \ref{ln:FTv} of Algorithm \ref{alg:main}. 
 
 \section{Concluding Remarks} \label{sec:conc}
We have proposed a fault localisation procedure for electrical networks possessing a location-parametrised admittance matrix. Its region of convergence must be a right half-plane, which need not contain the imaginary axis. Unstable networks are thereby accommodated, along with periodic or exponentially growing port signals.
A multiconductor transmission line with a voltage source at one end, sensors at the other, and a line-to-ground fault in the middle, is used as a case study. We show how to construct its admittance matrix from the admittances of its individual line segments, which are themselves modelled by the Telegrapher's equation. This choice of distributed-parameter model exemplifies infinite-dimensional dynamics with propagation delays, so the resulting network admittance is neither rational nor causal. While not every LTI network has an admittance matrix, the proposed localisation algorithm is easily adapted to other network models (e.g., impedance, chain, ABCD or scattering matrices) by following the same general strategy. 
 
 Working in the frequency domain, linearity is exploited to solve for the currents and voltages at the fault port as a function of its position, given the known port signals. Depending on the users knowledge of the fault impedance, two bivariate cost functions are constructed to yield the true fault location and time as a global minimum. If the impedance is known, frequency domain estimates are converted to the time-domain via inverse Fourier transform, and an error signal is then integrated over time to obtain the cost $J$. If unknown, passivity of the fault is assumed instead, which imposes non-negativity constraints over the open right-half plane. The cost $\tilde{J}$ is then obtained by integrating constraint violations over frequency. A fault-time estimator is also proposed as a function of fault distance, to reduce the dimension of the resulting minimisation problems. However, continuity of the two cost functions may not then be preserved. Numerical experiments on the case study found minimisation of $J$ to consistently yield accurate location estimates, and this was observed for any known fault resistance, regardless of its value. In the absence of fault impedance information, minimisation of $\tilde{J}$ produced accurate results for resistances below $5\ \Omega$.
 
 The limitations of our method and its analysis indicate directions for future work. Firstly, measurement noise, and numerical errors due to sampling and windowing, are yet to be considered. Since models of real-world large-scale networks are typically inaccurate, extending the analysis to include sources of both measurement and network model uncertainty is an important next step. In particular, very recent generalisations of passivity~\cite{moreschini2024generalized} offer a framework that can rigorously account for the sampling step. 
 Second, continuity of the two bivariate cost functions is only guaranteed if the known port signals are sufficiently smooth. Since this is difficult to know \emph{a priori}, the identification of practical sufficient conditions for Assumption~\ref{ass:smoothness} would in turn ensure a continuous minimisation problem at the end. Finally, the use of \emph{power transmission matrices}~\cite{aljanaideh_transmission_2019} enables the construction of admittances for other types of physical networks, such as water, gas, and even mechanical. Thus, our fault localisation methodology can potentially be adapted to these cases. Relaxing the linearity assumption would broaden the domain of applicability even further.
  
\printbibliography

\begin{IEEEbiography}[{\includegraphics[width=1in,height=1.25in,clip,keepaspectratio]{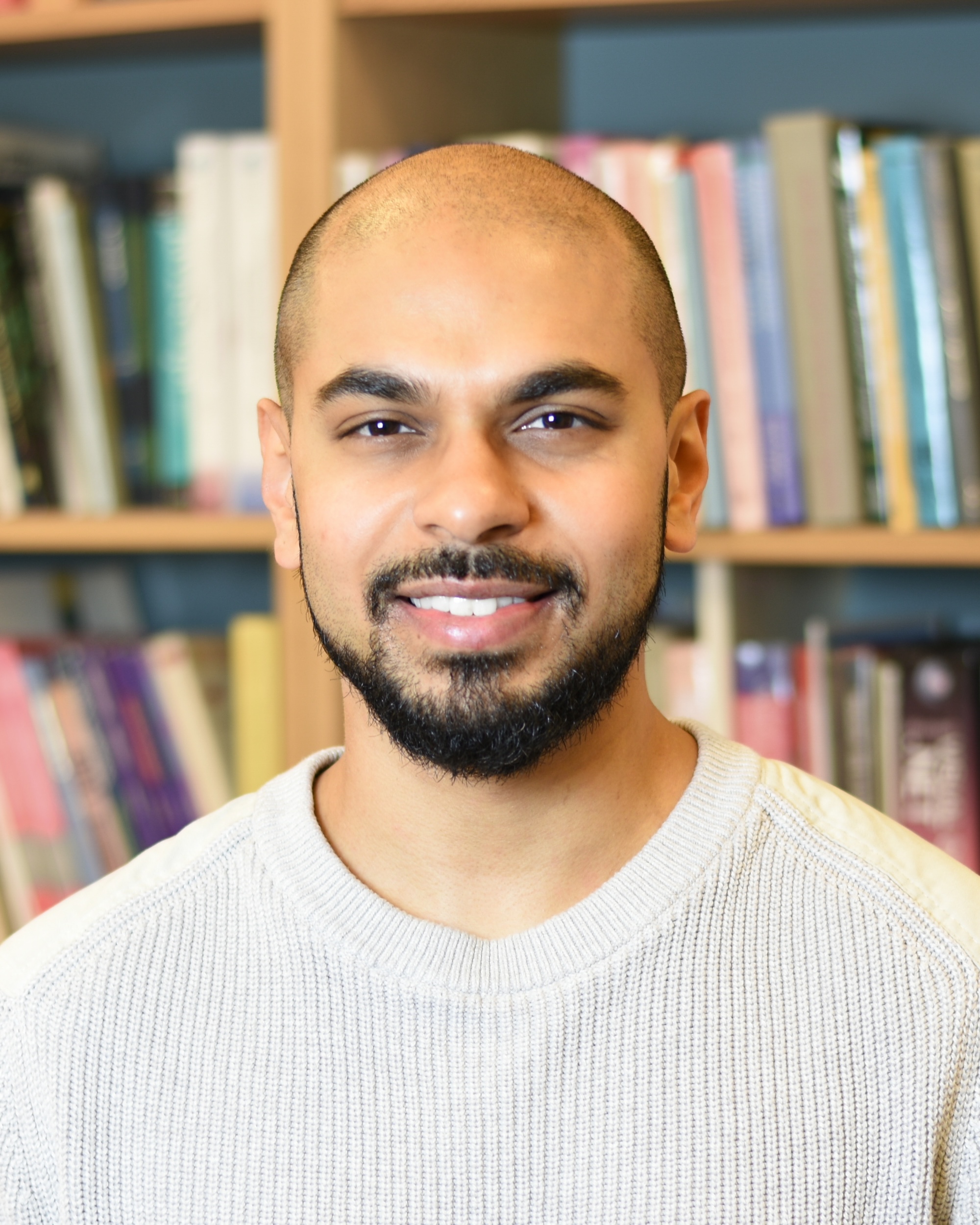}}]{Daniel Selvaratnam} (Member, IEEE) completed his B.Eng. in aerospace engineering with honours from Monash University, VIC, Australia, followed by his Ph.D. in electrical and electronic engineering from The University of Melbourne, VIC, Australia, in 2014 and 2019, respectively.
	
 After his Ph.D., he worked as Algorithm Engineer for BAE Systems Australia, on the guidance, navigation and control of autonomous aircraft. He was then Postdoctoral Research Fellow for The University of Melbourne, and is currently Postdoc at KTH Royal Institute of Technology in Stockholm. He has also been a visiting scholar at Imperial College, London, Lund University, Sweden, and Austrian Institute of Technology, Vienna. His research interests include control, estimation, and formal methods.
 
 Dr. Selvaratnam is a member of the Control Systems Society, and a professional member of Engineers Australia.
\end{IEEEbiography}

\begin{IEEEbiography}[{\includegraphics[width=1in,height=1.25in,clip,keepaspectratio]{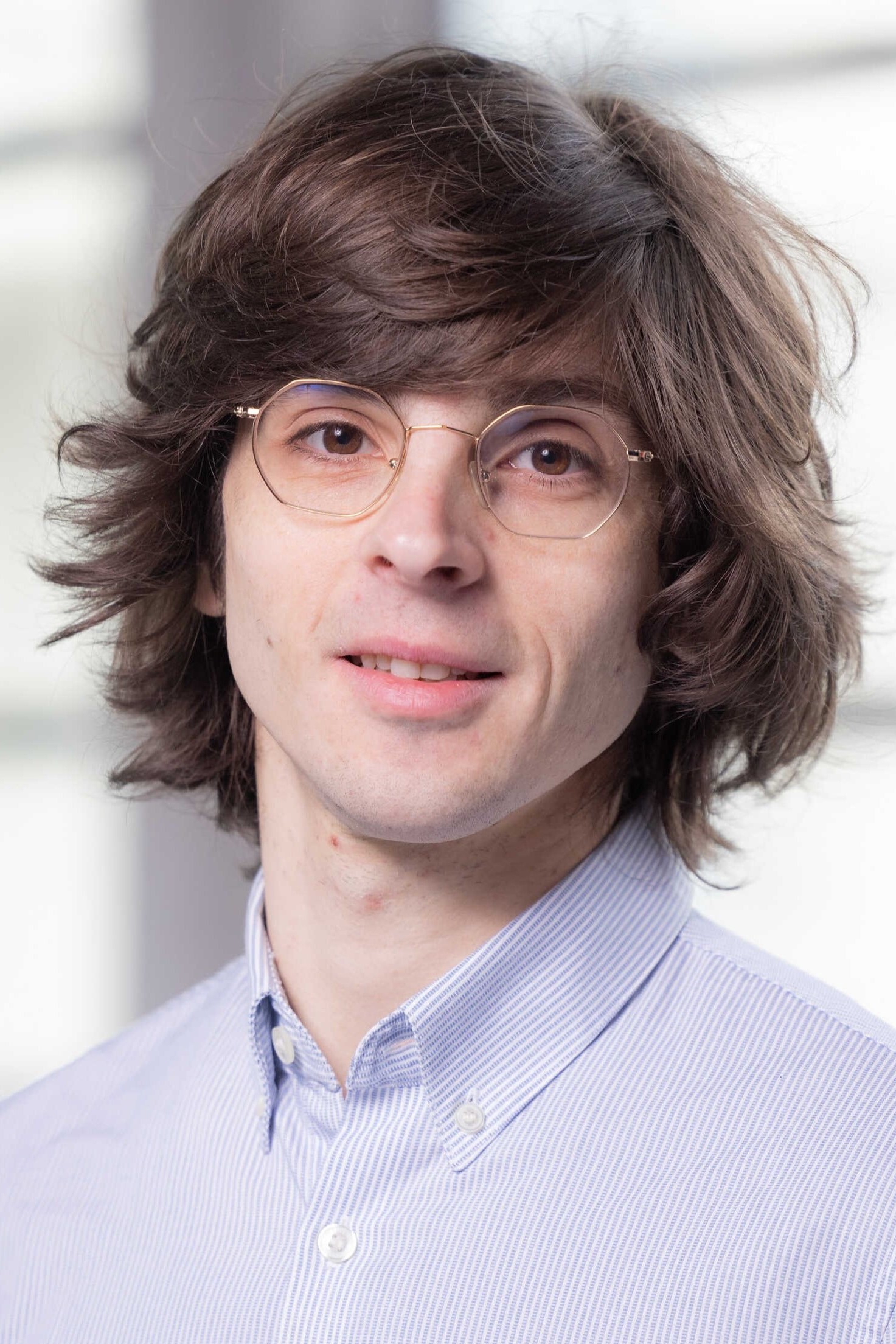}}]{Alessio Moreschini} (Member, IEEE) received the Ph.D. (Hons.) in Automatica from the University of Rome "La Sapienza," Italy and the Ph.D. in Systems and Control from Université Paris-Saclay, France, both in 2021. From 2021 to 2022, he was a Postdoctoral Fellow in the Department of Computer, Control, and Management Engineering at the University of Rome "La Sapienza." Since 2022, he has been a Research Associate in the Department of Electrical and Electronic Engineering at Imperial College London, UK. He serves as an Associate Editor for Automatica and the EUCA Conference Editorial Board. He is also a member of the IEEE CSS Technical Committee on Nonlinear Systems and Control and a Corresponding Member of the IFAC Technical Committee on Nonlinear Control Systems.
His research interests are in the field of nonlinear systems and control theory, with a particular focus on passivity-based control, sampled-data systems, and model order reduction.
\end{IEEEbiography}

\begin{IEEEbiography}[{\includegraphics[width=1in,height=1.25in,clip,keepaspectratio]{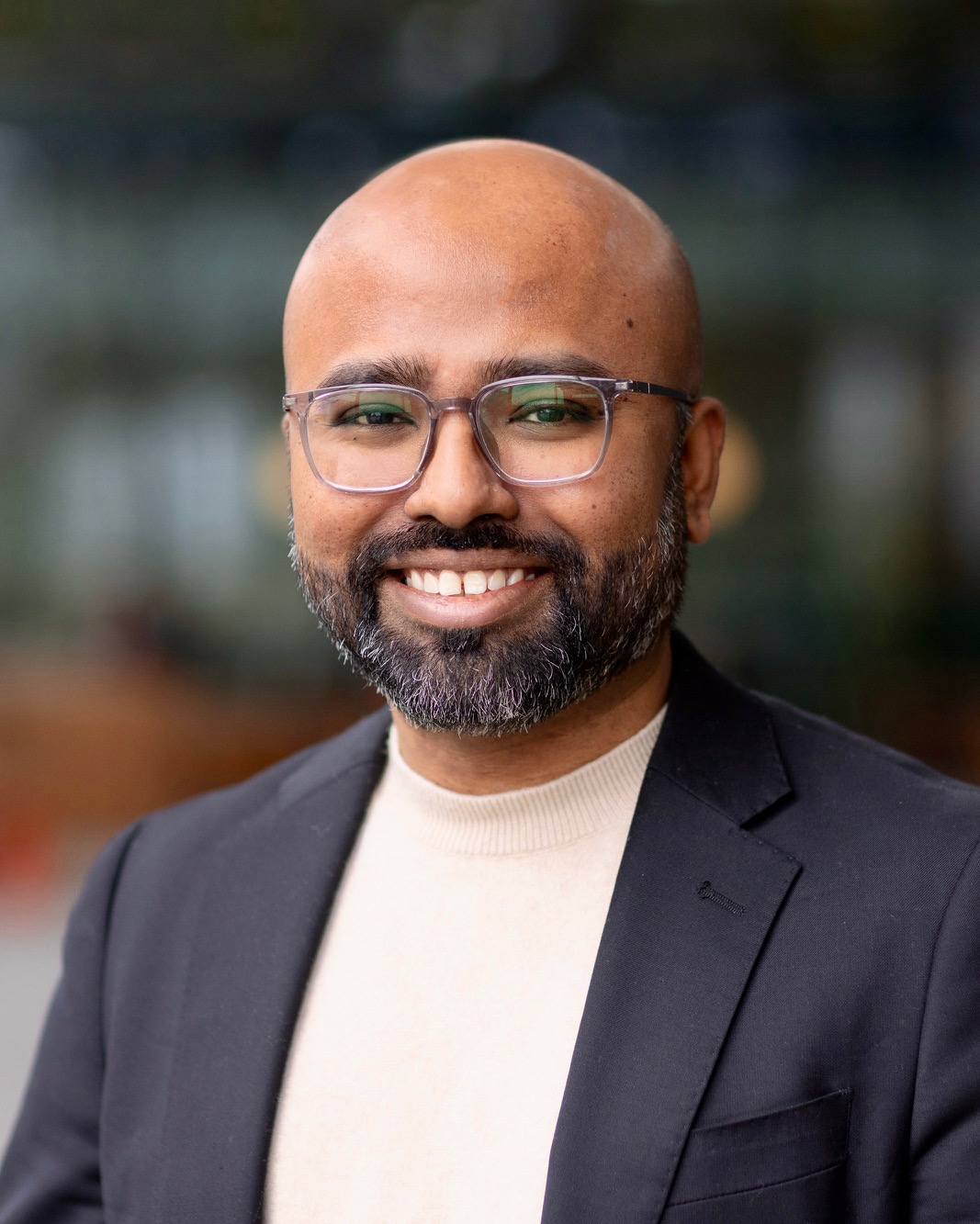}}]{Amritam Das} (Member, IEEE) received MSc. degree (2016) in Systems and Control and Ph.D. degree (2020) in Electrical Engineering from Eindhoven University of Technology. He held the position of research associate at the the University of Cambridge (2020-2021) where he was affiliated with Sideny Sussex College as a college research associate. During 2021-2023, he was a post-doctoral researcher at KTH Royal Institute of Technology. Since February 2023, He is an assistant professor at the Control Systems group of Eindhoven University of Technology. He currently serves as a member of IEEE CSS conference editorial board. He is also a member of IEEE/IFAC Technical Committee on Distributed Parameter Systems. His research interests are nonlinear control, physics-informed learning, control of PDEs, and model reduction. 
\end{IEEEbiography}

\begin{IEEEbiography}[{\includegraphics[width=1in,height=1.25in,clip,keepaspectratio]{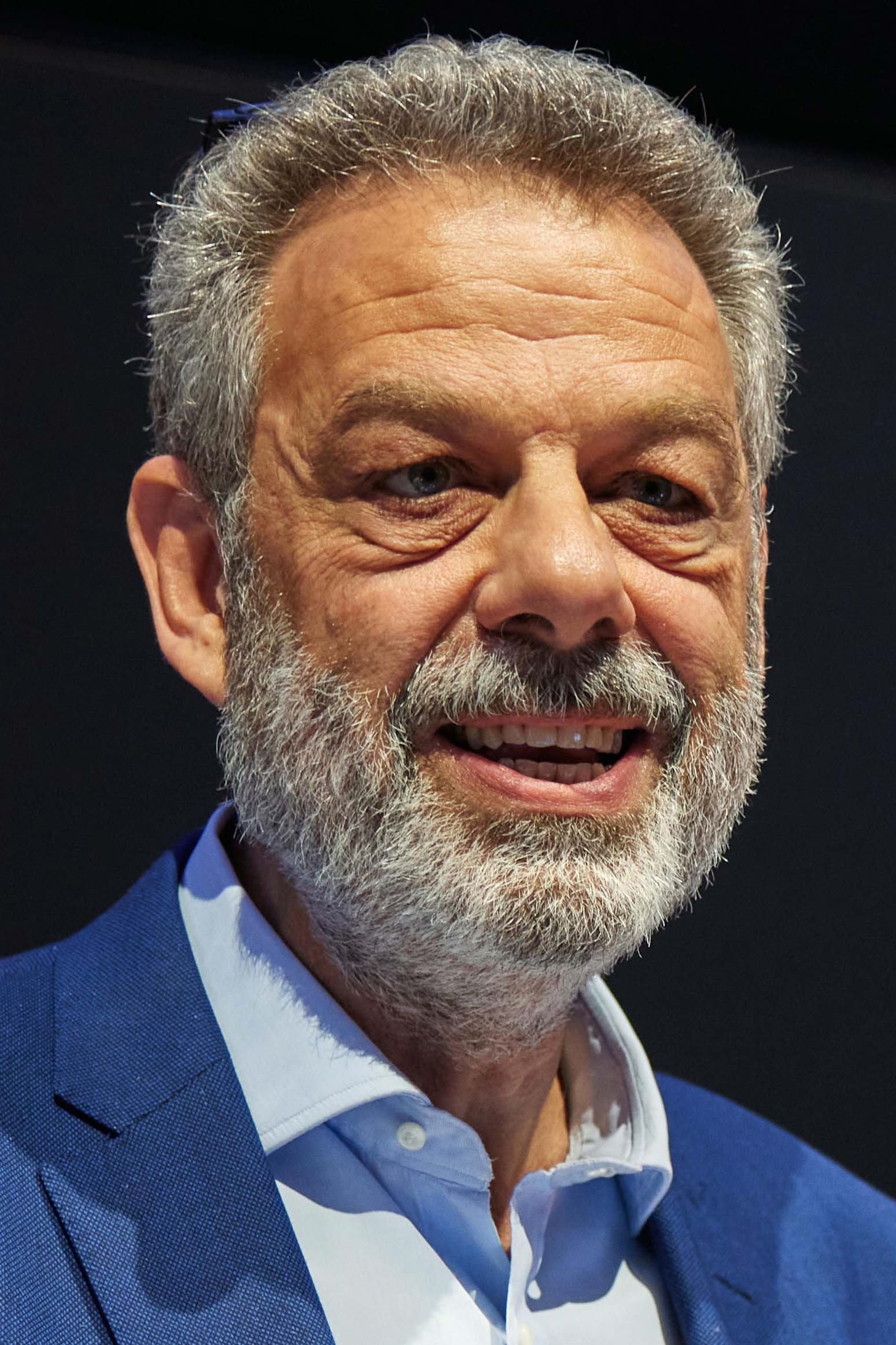}}]{Thomas Parisini} (Fellow, IEEE) received the Ph.D. degree in electronic engineering and computer science from the University of Genoa, Italy, in 1993. He was an Associate Professor with Politecnico di Milano, Milano, Italy. He currently holds the Chair of industrial control and is the Head of the Control and Power Research Group, Imperial College London, London, U.K. He also holds a Distinguished Professorship at Aalborg University, Denmark. Since 2001, he has been the Danieli Endowed Chair of automation engineering with the University of Trieste, Trieste, Italy, where from 2009 to 2012, he was the Deputy Rector. In 2023, he held a “Scholar-in-Residence”visiting position with Digital Futures-KTH, Stockholm, Sweden. He has authored or coauthored a research monograph in the Communication and Control Series, Springer Nature, and more than 400 research papers in archival journals, book chapters, and international conference proceedings. Dr. Parisini was the recipient of the Knighthood of the Order of Merit of the Italian Republic for scientific achievements abroad awarded by the Italian President of the Republic in 2023. In 2018 he received the Honorary Doctorate from the University of Aalborg, Denmark and in 2024, the IEEE CSS Transition to Practice Award. Moreover, he was awarded the 2007 IEEE Distinguished Member Award, and was co-recipient of the IFAC Best Application Paper Prize of the Journal of Process Control, Elsevier, for the three-year period 2011-2013 and of the 2004 Outstanding Paper Award of IEEE TRANSACTIONS ON NEURAL NETWORKS. In 2016, he was awarded as Principal Investigator with Imperial of the H2020 European Union flagship Teaming Project KIOS Research and Innovation Centre of Excellence led by the University of Cyprus with an overall budget of over 40 million Euros. He was the 2021-2022 President of the IEEE Control Systems Society and he was the Editor-in-Chief of IEEE TRANSACTIONS ON CONTROL SYSTEMS TECHNOLOGY (2009-2016). He was the Chair of the IEEE CSS Conference Editorial Board (2013-2019). Also, he was the associate editor of several journals including the IEEE TRANSACTIONS ON AUTOMATIC CONTROL and the IEEE TRANSACTIONS ON NEURAL NETWORKS. He is currently an Editor of Automatica and the Editor-in-Chief of the European Journal of Control. He was the Program Chair of the 2008 IEEE Conference on Decision and Control and General Co-Chair of the 2013 IEEE Conference on Decision and Control. He is a Fellow of IFAC. He is a Member of IEEE TAB Periodicals Review and Advisory Committee.
\end{IEEEbiography}

\begin{IEEEbiography}[{\includegraphics[width=1in,height=1.25in,clip,keepaspectratio]{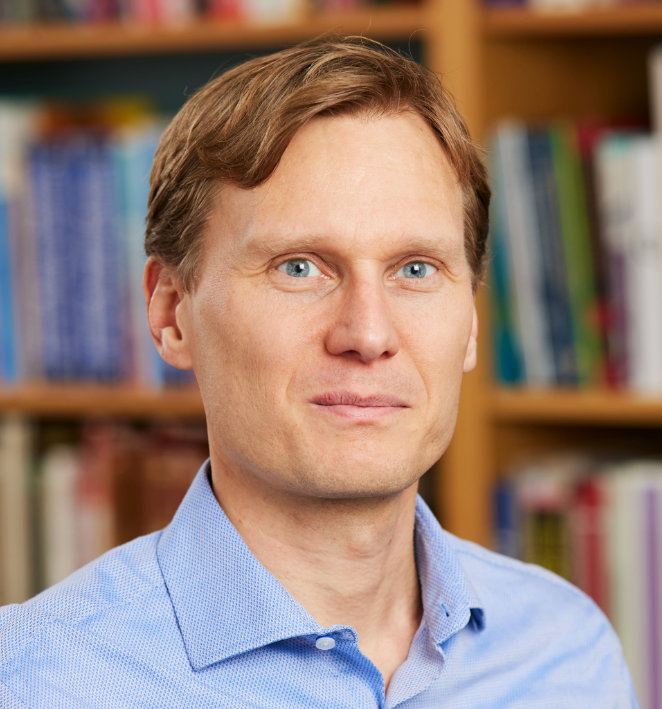}}]{Henrik Sandberg} (Fellow, IEEE) 
Henrik Sandberg is Professor at the Division of Decision and Control Systems, KTH Royal Institute of Technology, Stockholm, Sweden. He received the M.Sc. degree in engineering physics and the Ph.D. degree in automatic control from Lund University, Lund, Sweden, in 1999 and 2004, respectively. From 2005 to 2007, he was a Postdoctoral Scholar at the California Institute of Technology, Pasadena, USA. In 2013, he was a Visiting Scholar at the Laboratory for Information and Decision Systems (LIDS) at MIT, Cambridge, USA. He has also held visiting appointments at the Australian National University and the University of Melbourne, Australia. His current research interests include security of cyber-physical systems, power systems, model reduction, and fundamental limitations in control. Dr. Sandberg was a recipient of the Best Student Paper Award from the IEEE Conference on Decision and Control in 2004, an Ingvar Carlsson Award from the Swedish Foundation for Strategic Research in 2007, and a Consolidator Grant from the Swedish Research Council in 2016. He has served on the editorial boards of IEEE Transactions on Automatic Control and the IFAC Journal Automatica, and is currently an elected member of the IEEE Control Systems Society Board of Governors. He is Fellow of the IEEE.
\end{IEEEbiography}

\end{document}